\documentclass[%
reprint,
 amsmath,amssymb
 aps,
floatfix,
]{revtex4-2}

\usepackage{graphicx}% Include figure files
\usepackage{dcolumn}% Align table columns on decimal point
\usepackage{bm}% bold math
\usepackage{hyperref}% add hypertext capabilities
\usepackage{siunitx}[=v2]
\usepackage[version=4]{mhchem}
\usepackage{gensymb}
\usepackage{textpos}

\begin{document}

\title{Shell viscosity estimation of lipid-coated microbubbles}

% footnote(s) to article title
\homepage[Electronic Supplementary Information (ESI) available. See DOI: \phantomsection\label{lb:homepage} ]{https://doi.org/10.1039/d3sm00871a}

\author{Marco Cattaneo}
 \email{Electronic mail: mcattaneo@ethz.ch}
\author{Outi Supponen}%
\affiliation{%
 Institute of Fluid Dynamics, Department of Mechanical and Process Engineering, ETH Z{\"u}rich, Sonneggstrasse 3, 8092 Z{\"u}rich, Switzerland.
}%

\begin{abstract}
Understanding the shell rheology of ultrasound contrast agent microbubbles is vital for anticipating their bioeffects in clinical practice.
Past studies using sophisticated acoustic and optical techniques have made enormous progress in this direction, enabling the development of shell models that adequately reproduce the nonlinear behaviour of the coated microbubble under acoustic excitation.
However, there have also been puzzling discrepancies and missing physical explanations for the dependency of shell viscosity on the equilibrium bubble radius, which demands further experimental investigations.
In this study, we aim to unravel the cause of such behaviour by performing a refined characterisation of the shell viscosity.
We use ultra-high-speed microscopy imaging, optical trapping and wide-field fluorescence to accurately record the individual microbubble response upon ultrasound driving across a range of bubble sizes.
An advanced model of bubble dynamics is validated and employed to infer the shell viscosity of single bubbles from their radial time evolution.
The resulting values reveal a prominent variability of the shell viscosity of about an order of magnitude and no dependency on the bubble size, which is contrary to previous studies.
We find that the method called bubble spectroscopy, which has been used extensively in the past to determine the shell viscosity, is highly sensitive to methodology inaccuracies, and we demonstrate through analytical arguments that the previously reported unphysical trends are an artifact of these biases.
We also show the importance of correct bubble sizing, as errors in this aspect can also lead to unphysical trends in shell viscosity, when estimated through a nonlinear fitting from the time response of the bubble.
\end{abstract}

\maketitle
\begin{textblock*}{15cm}(0cm,-11.6cm)
  {\footnotesize \noindent Manuscript published in Soft Matter, 2023, 19, 5925 - 5941. \\ DOI: \href{https://doi.org/10.1039/D3SM00871A}{https://doi.org/10.1039/D3SM00871A}}
\end{textblock*}

\section{Introduction}

During the past decades, the development of stabilised gas microbubbles smaller than a red blood cell ($<\SI{10} {\micro\meter}$), and thus able to navigate the thinnest capillaries, has opened unique opportunities to utilise the  effects of ultrasonic cavitation in medicine.
Upon high-amplitude ultrasound driving, the microbubbles are set into a periodic succession of growth and collapse phases.
The microbubbles thus pump energy out of the acoustic wave and release it locally in the form of sound (even at frequencies other than the driving frequency) and mechanical action.
This extraordinary energy-focusing ability makes a simple bubble an extremely capable device for a large number of biomedical applications.
Coated microbubbles are currently used as ultrasound contrast agents to improve the contrast in perfusion imaging of the myocardium \cite{Wei1998QuantificationInfusion,Vogel2005TheValidation}, liver \cite{Wilson2000HarmonicMetastasis}, brain \cite{Rim2001QuantificationUltrasound}, kidney \cite{Wei2001QuantificationUltrasound} and other organs, providing unique information on blood volume and velocity owing to their large echogenicity.
The imaging potential of microbubbles can be augmented by equipping the coating with specific ligands.
Site-targeted contrast agents can be used to non-invasively image molecular events \emph{in vivo} such as
inflammation \cite{Lindner2001UltrasoundP-selectin,Kaufmann2007MolecularMolecule-1}, angiogenesis \cite{Leong-Poi2003Noninvasiveinfv/inf-integrins,Ellegala2003Imaginginfv/infinf3/inf} and early tumour formation \cite{Smeenge2017First-in-humanStudy,Willmann2017UltrasoundResults}. 
Moreover, microbubbles are presently investigated for therapeutic applications.
The high-energy mechanical effects of bubble cavitation (e.g. acoustic microstreaming \cite{Marmottant2003ControlledBubbles} and microjetting \cite{ Prentice2005MembraneCavitation}) can be harnessed to promote clot lysis in acute ischemic stroke \cite{Alexandrov2004Ultrasound-EnhancedTrial,Molina2006MicrobubbleActivator}, deliver plasmid DNA and drugs in the immediate perivascular region for myocardial infarction \cite{Qian2018TheTherapy}, atherosclerosis \cite{Yuan2018UltrasoundModel} and tumours treatment \cite{Dimcevski2016ACancer,Wang2018ClinicalSystem} or to locally permeabilise the blood-brain barrier to enhance the delivery of therapeutic agents for brain cancer \cite{Carpentier2016ClinicalUltrasound,Idbaih2019SafetyGlioblastoma,Mainprize2019Blood-BrainStudy}, Alzheimer’s disease \cite{Lipsman2018BloodbrainUltrasound} and amyotrophic lateral sclerosis \cite{Abrahao2019First-in-humanUltrasound} treatment.

A prolonged intravascular stability against dissolution and coalescence is achieved by encapsulating the microbubbles with a protein, biopolymer or lipid shell and by using a high-molecular-mass gas content such as perfluorocarbons and sulphur hexafluoride, which are less prone to outward loss compared to air \cite{Ferrara2007UltrasoundDelivery}.
To date, most clinically approved ultrasound contrast agents are coated with a phospholipid shell which is more flexible than other types of coatings, yielding less stable but more echogenic microbubbles.
Lipids naturally adsorb on gas-liquid interfaces and self-assemble into a monolayer which lowers the gas-liquid surface tension and thus the large Laplace pressure, arresting the gas efflux.
Moreover, it imparts rheological properties to the surface \cite{Edwards1991InterfacialRheology}.
Monolayers of DPPC (1,2-distearoyl-sn-glycero-3-phosphocholine), a frequent main lipid constituent of the bubble shell, show at low harmonic driving frequencies a dominantly viscous behaviour similar to three-dimensional liquids.
On the other hand, at higher frequencies their microstructure dominates the interfacial rheology and induces  an additional elastic contribution \cite{Hermans2014InterfacialConditions}.
This elasticity is of intrinsic rheological origin and is added to the Gibbs (apparent) monolayer elasticity, which arises when interface deformations yield variations of surfactant concentration \cite{Manikantan2020SurfactantFlows}.
The properties of the shell have a fundamental impact on the dynamical behaviour of ultrasound contrast agent microbubbles, and consequently, on their biophysical effects.
Compared to uncoated bubbles, coated bubbles show a higher resonance frequency due to a stiffer interface, a lower amplitude of response owing to viscous dissipation mechanisms between lipid molecules and a plethora of nonlinear effects \cite{Dollet2019BubbleMatter}.

Unfortunately, characterising the shell of a microbubble under clinically-relevant working conditions poses significant challenges owing to the micrometric size and the extremely high driving frequencies ($\sim \SI[parse-numbers=false]{10^6} {\hertz}$) to which these bubbles are subjected.
Direct experimental rheological techniques such as Langmuir trough and capillary pressure tensiometry methods are limited to significantly lower driving frequencies ($\sim \SI[parse-numbers=false]{1} {\hertz}$ and $\sim \SI[parse-numbers=false]{10^2} {\hertz}$, respectively)\cite{Derkach2009MethodsRheology} and at least one order of magnitude smaller film curvatures \cite{Alvarez2010AInterface}.
Therefore, as current direct approaches are not adequate for characterising the shell properties, only indirect methods can be used.
The seminal works of de Jong \textit{et al}. \cite{deJong1992AbsorptionMeasurements}, Frinking \& de Jong \cite{Frinking1998AcousticBubbles},  Hoff  \textit{et al}. \cite{Hoff2000OscillationsShell} and Gorce  \textit{et al}. \cite{Gorce2000InfluenceSonovueTM} provided the first estimates of the elastic and viscous contribution of the shell by performing acoustic attenuation measurements of polydisperse suspensions of coated microbubbles.
This approach was later refined by Parrales \textit{et al}.\cite{Parrales2014AcousticProperties} and Segers \textit{et al}. \cite{Segers2018High-precisionMicrobubbles} by employing monodisperse bubble suspensions produced by microfluidic flow–focusing techniques.
Moreover, Helfield \textit{et al.} \cite{Helfield2013NonlinearSpectroscopy} estimated the shell properties from scattered pressure measurements of single bubbles.
Morgan \textit{et al}. \cite{Morgan2000ExperimentalSize} provided the first optical characterisation by fitting a shell model to the radial evolution, captured on a streak camera image, of single ultrasound-driven microbubbles.
The same system was later also used by Doinikov \textit{et al.} \cite{Doinikov2009ModelingMicrobubbles}.
Only with the development of the groundbreaking Brandaris camera \cite{Chin2003BrandarisFrames} was the curtain on the wide-field time-resolved dynamics of microbubble finally lifted.
van der Meer \textit{et al}. \cite{VanDerMeer2007MicrobubbleAgents} and Overvelde \textit{et al}. \cite{Overvelde2010NonlinearMicrobubbles} were the first to leverage this new potential to characterise the shell of single DSPC-based microbubbles, followed by Luan \textit{et al.}\cite{Luan2012AcousticalMicrobubbles} to investigate the effect of liposomes loaded on the shell and van Rooij \textit{et al}.\cite{vanRooij2015Non-linearDPPC} to evaluate the influence of the carbon chain length of the phospholipids on the shell properties.
A side laser scattering system was used by Tu \textit{et al.} \cite{Tu2009EstimatingScattering,Tu2011MicrobubbleCytometry} to characterise the shell properties of commercial phospholipid-coated microbubbles by measuring their response to an ultrasound pulse.
The sensitivity of this technique has been further improved by employing a forward laser-scattering system that enabled Lum \textit{et al.} \cite{Lum2016SingleOscillations} to measure the decaying free oscillations and Lum \textit{et al.} \cite{Lum2018PhotoacousticProperties}, Supponen \textit{et al.} \cite{Supponen2020ThePopulations} and Daeichin \textit{et al.} \cite{Daeichin2021PhotoacousticAgents} to retrieve the frequency response of single bubbles set into small-amplitude oscillations via photoacoustic forcing.
Such small oscillations (few tens of nanometers) simplify the characterisation, as the absence of nonlinear effects allows the use of linear models.

In contrast to the dilatational modulus, for which there is broad empirical evidence regarding its independence from the bubble size, the reported dilatational viscosity values have been shown to increase with the bubble equilibrium radius and with a growing rate that strongly varies from study to study 
\cite{Morgan2000ExperimentalSize,VanDerMeer2007MicrobubbleAgents, Doinikov2009ModelingMicrobubbles,Tu2009EstimatingScattering, Tu2011MicrobubbleCytometry, Luan2012AcousticalMicrobubbles, Helfield2013NonlinearSpectroscopy,Parrales2014AcousticProperties,vanRooij2015Non-linearDPPC,Segers2018High-precisionMicrobubbles,Supponen2020ThePopulations,Daeichin2021PhotoacousticAgents}.
This observed dependency is widely regarded as unphysical, as it is illogical for a material property to be dependent on the quantity of the material.
van der Meer \textit{et al}. \cite{VanDerMeer2007MicrobubbleAgents}, Doinikov \textit{et al}. \cite{Doinikov2009ModelingMicrobubbles} and Tu \textit{et al}. \cite{Tu2011MicrobubbleCytometry} also showed the shell viscosity to decrease with the (maximum) strain rate of the bubble, suggesting a strain-thinning behaviour to possibly explain its questionable dependency on the bubble size.
However, this conclusion contradicts the observations from Helfield \textit{et al}. \cite{Helfield2013NonlinearSpectroscopy} and van Rooij \textit{et al}. \cite{vanRooij2015Non-linearDPPC} who found no significant relationship between shell viscosity and strain rate.
Moreover, the results reported by Tu \textit{et al}. \cite{Tu2011MicrobubbleCytometry} are particularly puzzling: as the parameters constituting the maximum strain rate $(\dot R / R)_{\rm max} \approx 2\pi f \Delta R_{\rm max} / R_0$ --- bubble resting radius $R_0$, maximum radial expansion (through pressure) $\Delta R_{\rm max}$ and driving frequency $f$ --- vary, the shell viscosity gives rise to different families of curves instead of collapsing on the same universal curve.
In our view, this means that the dependence of shell viscosity on the initial radius cannot be explained by a dependence on the strain rate.
Furthermore, the assumption that the shell viscosity is dependent on the strain rate throughout the entire radial excursion of the bubble pursuant to the Cross model, as proposed by Doinikov \textit{et al.} \cite{Doinikov2009ModelingMicrobubbles} following the result from van der Meer \textit{et al.}, is questionable.
In fact, during the expansion phase of the bubble, the phospholipids separate from each other, leading to a loss of the microstructure responsible for the non-Newtonian characteristics of the shell beyond a certain level of interfacial strain.
Conversely, during the compression phase of the bubble, the in-plane strain of the microstructure may have a minor role compared to the out-of-plane buckling.
Most studies reporting a dependency of shell viscosity on the bubble size have estimated the shell properties by analysing the bubble response in the frequency domain, adopting the "bubble spectroscopy" technique introduced by van der Meer \textit{et al}. \cite{VanDerMeer2007MicrobubbleAgents} or a variation of it.
The remaining works instead inferred the shell properties by fitting the experimental bubble radius-time curves.

In this work, we aim to clarify the apparent dependency of the shell viscosity on the microbubble size and the unexplained discrepancies in the values reported in different studies for similar bubbles.
We use ultra-high-speed microscopic imaging and optical trapping to directly observe the response of single ultrasonically driven coated microbubbles in an unbound fluid environment across a range of bubble radii.
We characterise the shell dilatational viscosity examining the bubble response in the time domain, instead of adopting the common bubble spectroscopy approach.
Contrary to other studies which fitted the bubble time response, we first confronted our experimental data against the theoretical predictions to verify that the chosen bubble model is capable of describing the salient features of the bubble response, in particular the bubble resonance size.
The shell dilatational viscosity is then inferred by fitting the experimental response with the theoretical one.
Finally, the results are compared with previous studies and an explanation for the origin of the unphysical dependence of shell viscosity on the microbubble size is provided.

\section{Theory}\label{BubbleModel}
The radial motion of a spherical bubble of radius $R(t)$ is described by a modified form of the Rayleigh--Plesset equation for mildly compressible Newtonian media \cite{Prosperetti1986BubbleTheory,Lezzi1987BubbleTheory,Brenner2002Single-bubbleSonoluminescence,Prosperetti2013AOscillations},  which includes a pressure term $ \Sigma\bigl(R, \dot R\bigr)$ to account for the generalised interfacial stresses $\boldsymbol{\sigma}_s \bigl(R,\dot R \bigr)$ and reads
\begin{multline}\label{eq:RP}
\rho_l \left(R \ddot R + \frac{3}{2} \dot R^2 \right)= \\ \left( 1 + \frac{R}{c_l} \frac{d}{dt} \right) p_g
   +  \Sigma\bigl(R, \dot R\bigr) - p_{\infty} - p_d\left(t\right) - 4\mu_l \frac{\dot R}{R},
\end{multline}
where over-dots denote time differentiation, $\rho_l$ the liquid density, $c_l$ the speed of sound in the medium, $p_g$ the gas pressure inside the bubble, $p_{\infty}$ the undisturbed ambient pressure, $p_d(t)$ the acoustic driving pressure and $\mu_l$ the dynamic viscosity of the medium.

The pressure contribution of the interface $\Sigma\bigl(R, \dot R\bigr)$ is given by the radial component of the surface divergence of the interfacial stress \cite{Edwards1991InterfacialRheology,Dollet2019BubbleMatter}
\begin{equation}\label{eq:InterfacePressure}
 \Sigma\bigl(R, \dot R\bigr) = \left[\nabla_s \cdot \boldsymbol{\sigma}_s \bigl(R,\dot R \bigr)\right]_r,
\end{equation}
where $\nabla_s  \stackrel{\rm def}{=} \boldsymbol{I}_s \cdot \nabla$ is the surface gradient operator, $\boldsymbol{I}_s \stackrel{\rm def}{=} \boldsymbol{I} - {\boldsymbol{n}}{\boldsymbol{n}}$ is the surface unit tensor and ${\boldsymbol{n}}$ is the unit vector normal to the interface.
As opposed to earlier models that treated the shell as a three-dimensional solid \cite{Church1995TheBubbles,Hoff2000OscillationsShell,Morgan2000ExperimentalSize}, Chatterjee and Sarkar \cite{Chatterjee2003AAgents,Sarkar2005CharacterizationEncapsulation} were the first to suggest modelling the bubble coating as a two-dimensional continuum. This approach is considered the most appropriate for molecular membranes (i.e.\ phospholipid \cite{Borden2008AImaging} and protein \cite{Christiansen1994PhysicalAlbumin} shells) for which the assumption of continuum in the normal direction is questionable \cite{Pozrikidis2003ModelingCells}.
They proposed to model the microbubble shell as a viscoelastic solid interface, expressing the surface stress tensor, owing to the spherical symmetry of the problem, as
\begin{equation}\label{eq:InterfaceStressTensor}
    \boldsymbol{\sigma}_s \bigl(R,\dot R \bigr) = \left[\sigma_0 + E_s\left(J-1\right) \right]\boldsymbol{I}_s + \kappa_s \bigl(\nabla_s \cdot \dot R \ {\boldsymbol{n}}\bigr) \boldsymbol{I}_s.
\end{equation}
The first term represents the isotropic part of the Evans--Skalak model for elastic interfaces \cite{Evans1980MechanicsBiomembranes} without differentiating 
the thermodynamic contribution from the rheological one \cite{Jaensson2018TensiometryInterfaces,Jaensson2021ComputationalRheology} and the second the isotropic part of the Boussinesq--Scriven model for viscous interfaces \cite{Edwards1991InterfacialRheology} which is the surface equivalent of the Newtonian model for compressible bulk fluids.
$\sigma_0$ is the interfacial surface tension at the equilibrium bubble radius $R_0$, $E_s$ is the interfacial dilatational modulus, $\kappa_s$ is the interfacial dilatational viscosity and  $J=R^2 / R_0^2$ is the relative area deformation. 
Upon substitution of Eq.\ \eqref{eq:InterfaceStressTensor} into Eq. \eqref{eq:InterfacePressure} and by making use of the following differential identities, valid for a spherically symmetric problem, $\nabla_s \cdot \boldsymbol{n} = {2}/{R}$ and $\nabla_s \cdot \boldsymbol{I}_s = -\left({2}/{R}\right) \boldsymbol{n}$, the pressure contribution given by the interface reads
\begin{equation}\label{eq:InterfacePressureSarkar}
 \Sigma\bigl(R, \dot R\bigr) = - 2\frac{\sigma_0 + E_s \left(J-1\right)}{R} -4\kappa_s\frac{\dot R}{R^2}.
\end{equation}
To account for the strong nonlinear effects that coated microbubbles exhibit, such as subharmonic oscillations \cite{Lotsberg1996ExperimentalBubbles,Shankar1998AdvantagesEnhancement,Shankar1999SubharmonicAgents,Sun2005High-frequencyAgents,Sijl2010SubharmonicMicrobubbles}, compression-only behaviour \cite{Marmottant2005ARupture}, thresholding behaviour \cite{Emmer2007TheVibration} and resonance frequency shifting \cite{Overvelde2010NonlinearMicrobubbles}, which a quasi-linear shell model as Eq.\ \eqref{eq:InterfacePressureSarkar} cannot represent, Marmottant \textit{et al}. \cite{Marmottant2005ARupture} proposed a phenomenological extension inspired by the behaviour of liquid-gas interfaces covered by insoluble surfactants \cite{Hermans2014InterfacialConditions}, yielding
\begin{equation}\label{eq:InterfacePressureMarmottant}
\resizebox{8.6cm}{!}{$
      \Sigma\bigl(R, \dot R\bigr) = 
\begin{cases}
 -4\kappa_s\displaystyle\frac{\dot R}{R^2},  \! \! \! \! & \text{for } R\leq R_{\rm b}\\
   - 2\displaystyle\frac{\sigma_0 + E_s \left(J-1\right)}{R} -4\kappa_s\displaystyle\frac{\dot R}{R^2},  \! \! \! \! & \text{for } R_{\rm b}< R < R_{\rm r}\\
    - 2\displaystyle\frac{\sigma_{\rm water}}{R}  -4\kappa_s\displaystyle\frac{\dot R}{R^2},  \! \! \! \! & \text{for } R \geq R_{\rm r}
\end{cases}$}
\end{equation}
which accounts for: (i) the buckling of the coating when the bubble is compressed below its shell compression limit radius $R_{\rm buckling}$ resulting in a tensionless interface, (ii) the rupturing of the coating when the bubble is expanded beyond $R_{\rm rupture}$ where it experiences a surface tension equal to that of a pure gas-water interface $\sigma_{\rm water} = \SI {72.8} {\milli\newton\per\meter}$, and (iii) the elastic regime in between.
The Marmottant model has proven to be a compelling engineering representation of the microbubble shell behaviour, capable of explaining a broad spectrum of nonlinearities \cite{Overvelde2010NonlinearMicrobubbles,Sijl2010SubharmonicMicrobubbles}, and to be sophisticated enough to accurately predict the behaviour of bubbles in the present study, without the need for additional free parameters.
However, this model is not entirely free from physical inconsistencies, an important one being the assumption of a constant shell viscosity for varying the free molecular area. Nevertheless, this can be justified as being an average viscosity over the radial excursion of the bubble.

In order to close the problem, the gas pressure contained in the bubble must be determined, in principle, by solving the full Navier--Stokes equation inside the bubble.
This can be done analytically only for small-amplitude oscillations for which the governing equations can be linearised \cite{Chapman1971ThermalBubbles,Prosperetti1977ThermalLiquids} or numerically at a high computational cost.
For these reasons, extensive use is made of the crude approximation of a polytropic process:
\begin{equation}\label{eq:PolyGas}
p_g = p_{g,0} \left(\frac{R_0}{R}\right)^{3n}, \ p_{g,0} = p_{\infty} + \frac{2\sigma_0}{R_0},
\end{equation}
with $p_{g,0}$ the bubble internal pressure at rest and $n$ the polytropic index.
In spite of its appealing simplicity, this approximation suffers from serious limitations \cite{Prosperetti1988NonlinearDynamics}. 
First, its range of validity is restricted to low-amplitude oscillations and to global Péclet numbers $\text{Pe}=R_0^2 \omega/ D_g$ (where $\omega = 2 \pi f$ is the driving angular frequency and $D_g$ is the gas thermal diffusivity) much smaller or much bigger than one, and therefore to the particular cases where the gas follows an isothermal process ($n = 1$) or an adiabatic one ($n = \gamma$, where $\gamma$ is the gas specific heat ratio), respectively, for the whole radial motion of the bubble. 
And second, its use results in no energy loss associated with the change of temperature of the gas.
These restrictions are unfortunate in the context of ultrasound contrast agent microbubbles because under clinical conditions they often undergo large nonlinear oscillations, characterised by a thermal motion defined by a global $\text{Pe}\sim 1$ with both isothermal and adiabatic behaviours occurring at different phases of oscillation, and subjected to a thermal loss of the same order of magnitude of the viscous damping.
To address the latter point, it is customary in the literature to artificially increase the liquid viscosity doubling its value \cite{Versluis2020UltrasoundReview}.
These considerations motivate us to take a step beyond the polytropic approximation.
In an attempt to simplify the problem, Nigmatulin \textit{et al.} \cite{Nigmatulin1981DynamicsLiquid} and Prosperetti \textit{et al.} \cite{Prosperetti1988NonlinearDynamics} independently found the spatial variation of the pressure inside the bubble to be negligible in most cases ($\Delta p_g / p_g \sim O( R \dot R / \lambda c_g, \dot R^2 / c_g^2)$, where $\lambda$ and $c_g$ are representative values of the sound wavelength and speed in the bubble), allowing, for a perfect gas, to exactly express the gas radial velocity as:
\begin{equation}\label{eq:ug}
u_g = \frac{1}{\gamma p_g}\left[\left(\gamma-1\right) K_g \frac{\partial T_g}{\partial r} -\frac{1}{3} r \dot p_g \right],
\end{equation}
and from this result recover an exact relationship for the gas pressure:
\begin{equation}\label{eq:pg}
\dot p_g = \frac{3}{R}\left[\left(\gamma-1\right) K_g \left.\frac{\partial T_g} {\partial r}\right|_R -\gamma p_g \dot R \right],
\end{equation}
where $r$ is the radial coordinate, $T_g$ is the gas temperature, and $K_g$ is the thermal conductivity of the gas.
These two relationships allow to reduce the number of governing partial differential equations (PDEs) from three to one, the one necessary to find the temperature gradient at the bubble surface, i.e.\ $\left. {\partial_r}T_g\right|_R$, which is the term that governs the heat exchange and thus closes the problem.
One can solve either the energy equation or the continuity equation coupled with the equation of state for perfect gases.
The first alternative was chosen by Kamath \& Prosperetti \cite{Kamath1989NumericalDynamics}, who solved the energy equation by using a Galerkin-Chebyshev spectral method.  
Recently, the second alternative was pursued by Zhou \cite{Zhou2021ModelingBubble}, who found the temperature profile inside the bubble to be divided into three regions: (1) an internal layer characterised by uniform temperature, (2) a buffer layer, and (3) a layer adjacent to the bubble surface characterised by a linear temperature distribution, which in turn suggests dividing the computational domain into as many cells.
In this way, the continuity equation 
\begin{equation}\label{eq:continuity}
\dot \rho_g + \nabla \cdot  \left( \rho_g \boldsymbol{u}_g \right) = 0,
\end{equation}
where $\rho_g$ denotes the gas density,
can be converted into two ordinary differential equations (ODEs)
\begin{equation}\label{eq:discretecontinuity}
\begin{cases}
\dot m_{g,1} = -f_1 = -\rho_{g,1}^{\rm uw}u_{g,1}^{\rm rel}S_1, \\
\dot m_{g,3} = f_2 = \rho_{g,2}^{\rm uw}u_{g,2}^{\rm rel}S_2,
\end{cases}
\end{equation}
where $m_{g,i}$ is the gas mass in cell $i$ and $f_j$ is the mass flux across the cell interface $j$.
$\rho_{g,j}^{\rm uw}$ is the density of the neighboring cell of interface $j$ in the upwind direction, $u_{g,j}^{\rm rel}$ is the convective velocity and $S_j$ is the interface $j$ surface area.
The velocity $u_g$ can be obtained from Eq.\ \eqref{eq:ug}, while the temperature $T_g$  computed through the equation of state for ideal gas
\begin{equation}\label{eq:idealgaslaw}
p_g = \rho_g\mathcal{R} T_g,
\end{equation}
where $\mathcal{R}$ is the the specific gas constant.
The change in bubble surface temperature is only a small fraction of that of the gas and can, therefore, be neglected, i.e. $T_g|_R \approx T_l$ \cite{Prosperetti1988NonlinearDynamics}.
At this point, the motion of the bubble can be effortlessly solved by integrating in time the system of ODEs consisting of Eq.\ \eqref{eq:RP}, \eqref{eq:pg} and \eqref{eq:discretecontinuity}.
The reader can refer to the article by Zhou \cite{Zhou2021ModelingBubble} for full details of the method.
The simplicity and accuracy of the model make it suitable for this work.

\section{Methods}

\subsection{Microbubble synthesis}
Lipid-coated microbubbles with a \ce{C4F10} (perfluorobutane, Fluoromed) gas core are prepared in-house by probe sonication. 
The lipid coating consists of 90 mol\% {DPPC} (1,2-distearoyl-sn-glycero-3-phosphocholine, NOF EUROPE) and 10 mol\%  of  {DPPE-PEG5K} (1,2-dipalmitoyl-sn-glycero-3-phosphoethanolamine-N-[methoxy(polyethylene glycol)-5000]), Avanti Polar Lipids).
The lipids are first dissolved in chloroform.
The solution is placed under vacuum at $\SI{35}{\celsius}$ overnight to remove the solvent and produce a dry lipid film. 
The dry lipid film is rehydrated with $1\times$ PBS (phosphate buffered saline, Boston BioProducts) to yield a total lipid concentration of $ \SI{2}{\milli\gram\per\milli\litre}$.
The solution is then sonicated (SFX550, Branson; $\SI{20} {\kilo\hertz}$, $\SI{550} {\watt}$) at low power (30\%) for five minutes to convert the multilamellar vesicles to unilamellar liposomes. 
Microbubbles are formed by probe-sonicating the surface of the lipid solution at full power for ten seconds while simultaneously flowing  \ce{C4F10} gas over it.
The microbubble suspension is then cooled down to room temperature and washed using centrifugation.
Differential centrifugation is used to size-isolate the microbubbles into a specific size distribution ($ \SIrange{1}{4}{\micro\metre}$-radius), based on the protocol developed by Feshitan \textit{et al}. \cite{Feshitan2009MicrobubbleCentrifugation}.
Fluorescent-labeled microbubbles are fabricated by adding the lipid dye DiI (1,1'-dioctadecyl-3,3,3'3'-tetramethylindocarbocyanine perchlorate, Sigma-Aldrich) to the lipid solution at a concentration of $ \SI{5}{\micro\gram\per\milli\litre}$ before probe sonication.

\subsection{Experimental setup}
\begin{figure}[ht!] 
    \centering
        \includegraphics[width=\columnwidth]{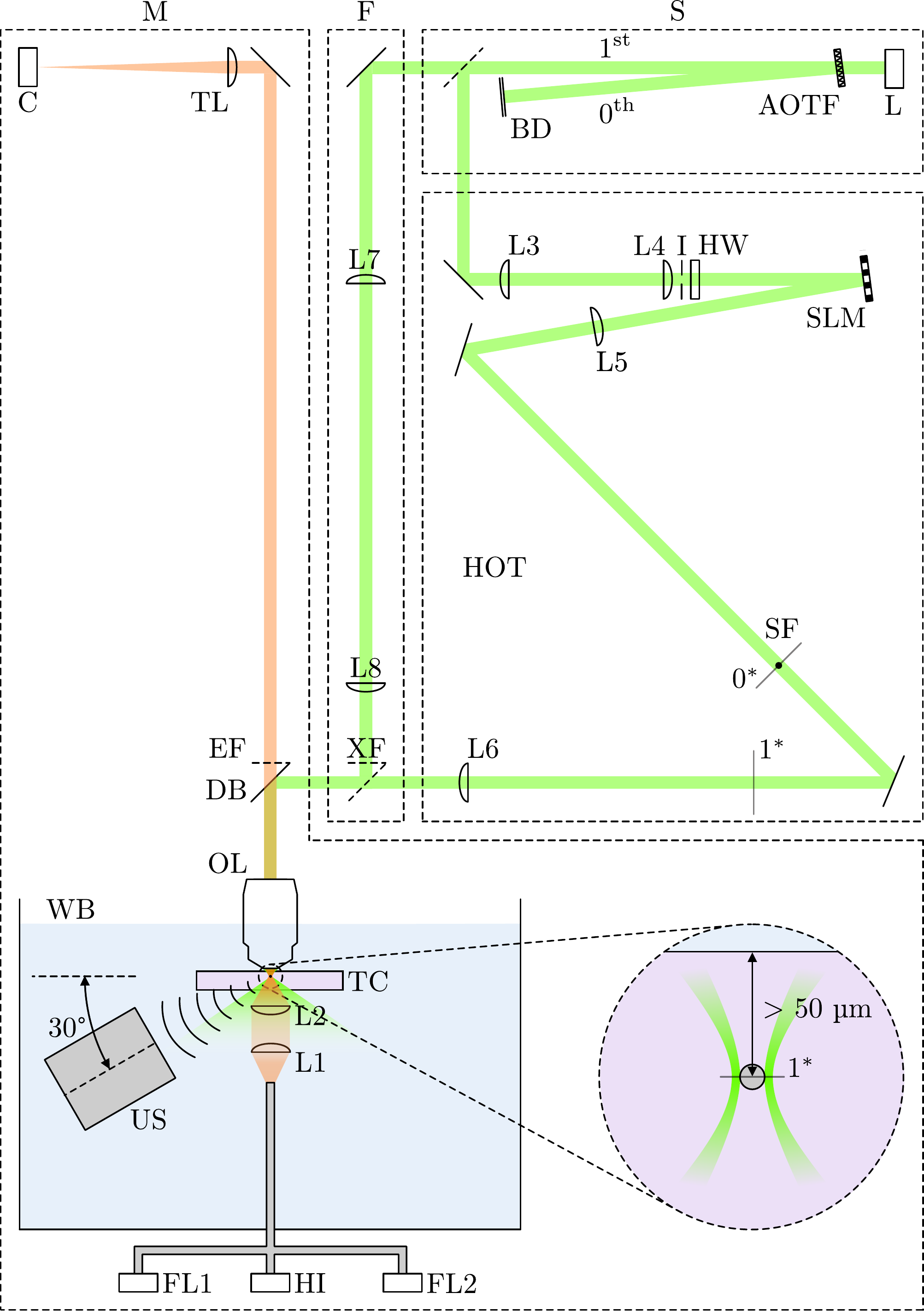}
    \caption{Schematics of the experimental setup. (M) Microscope subsystem, (F) Fluorescence subsystem, (S) Acousto-optic shutter subsystem, (HOT) Holographic optical tweezers subsystem. (AOTF) Acousto-optic tunable filter, (BD) Beam dump, (C) Camera, (DB) Dichroic beamsplitter, (EF) Emission filter, (FL1-FL2) Flash lamp, (HI) Halogen illuminator, (HW) Half-wave plate, (I) Iris, (L) Laser, (L1-L8) Lens, (OL) Objective lens, (SF) Spatial filter, (SLM) Spatial light modulator, (TC) Test chamber, (TL) Tube lens, (US) Ultrasound transducer, (WB) Water bath, (XF) Excitation filter. ($0^*$) Zeroth-order laser beam image plane, ($1^*$) First-order laser beam image plane. The zoomed-in inset depicts the pre-test conditions, with a single microbubble being optically trapped $> \SI{50} {\micro\meter}$ from the top window of the test chamber. }
    \label{fig:setup}
\end{figure}

To study the dynamics of microbubbles, we developed a multimodal microscope ("Mscope", Fig.\ \ref{fig:setup}) which allows for wide-field ultra-high-speed imaging, holographic optical trapping and wide-field fluorescence.

The main body of the system  [Fig.\ \ref{fig:setup}(M)] is a custom-built upright microscope realised using modular optomechanics components (Thorlabs, cage system), installed on an optical table  (T1220C, Thorlabs) with active isolators (PTS603, Thorlabs) to minimise environmental vibrations. 
The microscope is equipped with a water-dipping  objective lens of focal length  $\mathcal{F} = \SI{2} {\milli\meter}$ (CFI Plan 100XC W, Nikon) and a $\mathcal{F} = \SI{400} {\milli\meter}$ tube lens (TL400-A, Thorlabs) for a total magnification of $200\times$. 
An ultra-high-speed camera (HPV-X2, Shimadzu) allows for recordings at 10 million frames per second over $\SI{25.6}{\micro\second}$ of continuous visualisation of a $\SI{64 x 40} {\micro\meter}$ field of view with a 160-nm pixel resolution.
Backlight illumination is provided by a continuous halogen illuminator (OSL2, Thorlabs) for live imaging and two Xenon flash lamps used sequentially (MVS-7010, EG\&G) for video recording, all combined into a single optical fiber output and focused on the sample through a custom-built condenser (L1 and L2, $\mathcal{F} = \SI{25} {\milli\meter}$, AC127-025-A, Thorlabs).
A water bath $(T_l\approx \SI{22}{\celsius} )$ filled with deionised water is accommodated at the base of the microscope. 
A custom-designed 3D-printed test chamber with optically and acoustically transparent windows on top and bottom (low-density polyethylene, acoustic impedance $z=\SI{1.79e6}{\pascal\second\per\meter}$) is suspended in the water bath and its position controlled by a three-axis motorised microtranslation stage (PT3/M-Z8, Thorlabs).
A broadband focused ultrasound transducer (PA1612, Precision Acoustics; $\SI{1.5} {\mega\hertz}$ of center frequency, $\SI{85} {\milli\meter}$ of focal length, $\approx \SI{4} {\milli\meter}$ of beam width at $\SI{-6} {\deci\bel}$) is positioned in the water bath at an angle of $30 \degree$ with respect to the horizontal plane to minimise acoustic reflections and manoeuvred using a manual three-axis microtranslation stage (\#12694, \#66-511, Edmund Optics).
A function generator (LW 420B, Teledyne LeCroy) is used to produce the driving pulse, which is then amplified by a radiofrequency power amplifier (1020L, E\&I). 
A calibrated needle hydrophone ($\SI{0.2} {\milli\meter}$, NH0200, Precision Acoustics) is employed to measure the driving acoustic pressure inside the test chamber and to align the acoustic focal point with the optical field.

The holographic optical tweezers and fluorescent microscopy are realised using a 532-nm optically pumped semiconductor continuous-wave laser (Verdi G10, Coherent).
The laser beam enters an acousto-optic tunable filter (AOTF.NC-VIS/TN, AA Opto Electronic) that functions as a microsecond-fast electronic shutter [Fig.\ \ref{fig:setup}(S)].
The transmitted beam ($0^{\rm th}$ order) is ceased with a beam dump while the diffracted beam ($1^{\rm st}$ order), whose intensity is dependent upon the radiofrequency drive power provided by a radiofrequency driver (MOD.8C.10.b.VIS,  AA Opto Electronic), is selectively sent to the optical tweezers or the fluorescence setup by means of a movable mirror.

Three-dimensional optical trapping of single microbubbles has been demonstrated in the pioneering works of Prentice \textit{et al}. \cite{Prentice2004ManipulationArrays} and Garbin \textit{et al}. \cite{Garbin2005OpticalBeams} by focusing an optical vortex laser beam obtained by converting the laser transverse mode from Gaussian to Laguerre-Gaussian by means of a diffractive optical element.
Compared to the infrared laser employed in those studies, using a green laser reduces the local heating of water by three orders of magnitude \cite{Hale1973OpticalRegion,Pope1997AbsorptionMeasurements}.
In our implementation [Fig.\ \ref{fig:setup}(HOT)], the optical path of the optical tweezers setup is kept as short as possible to achieve the highest pointing stability.
The diffractive optical element (DOE) is provided by a reflective spatial light modulator (SLM) (PLUTO-2.1 VIS-096, HOLOEYE Photonics).
The SLM is placed after a $7.5\times$ galileian telescope, made of a pair of two lenses (L3, $\mathcal{F} = \SI{-20} {\milli\meter}$, ACN127-020-A, Thorlabs; L4, $\mathcal{F} = \SI{150} {\milli\meter}$, AC254-150-A, Thorlabs), and an iris (SM1D12, Thorlabs) which expand and spatially filter the laser beam to completely fill the active area of the SLM with an approximately uniform-intensity distribution which helps to generate a strong optical trap.
The linear polarisation of the laser is rotated by using a zero-order half-wave plate (WPHSM05-532, Thorlabs) to match the alignment of the liquid crystalline molecules of the SLM microdisplay.
The laser beam mode is converted into a Laguerre-Gaussian one by implementing an optical vortex phase mask on the SLM.
In order to minimise the diffraction effects, a $4\mathcal{F}$ system is used to relay the SLM image plane to the objective back focal plane (L5, $\mathcal{F} = \SI{400} {\milli\meter}$, AC254-400-A, Thorlabs; L6, $\mathcal{F} = \SI{200} {\milli\meter}$, AC254-200-A, Thorlabs).
The magnification of the $4\mathcal{F}$ system is $0.5\times$ in order to slightly overfill the objective lens back aperture and thus obtain the best optical trapping performance.
The $0^{\rm th}$-order component of the beam, unavoidably present due to the pixelated structure of the SLM, is suppressed by performing spatial filtering after having separated the $1^{\rm st}$-order-component image plane from the $0^{\rm th}$ one by superimposing a phase mask for a lens onto the SLM.
Upon reflection on a reflective-band dichroic beamsplitter (ZT532dcrb, Chroma), the beam is focused on the sample by the objective lens. 
The optical trap can be freely positioned within the sample image plane by superimposing a phase mask for a prism onto the SLM.

With the use of movable mirrors the optical tweezers setup can be bypassed and make use of the laser for performing epifluorescence microscopy [Fig.\ \ref{fig:setup}(F)]. 
The size of the field of illumination and consequently the laser-excitation power density can be adjusted varying the spacing of a lens relay system (L7 and L8, $\mathcal{F} = \SI{100} {\milli\meter}$, AC254-100-A, Thorlabs).
Undesirable light wavelengths are prevented from illuminating the specimen by means of a narrow passband excitation filter (ZET532/10x, Chroma). 
The dichroic beamsplitter reflects the excitation light into the objective lens. 
The laser line is removed from the specimen image with an emission filter (ET590/50m, Chroma).

\section{Results} \label{Results}
\subsection{Microbubble radius measurement}
\begin{figure}[!ht] 
    \centering
        \includegraphics[width=\columnwidth]{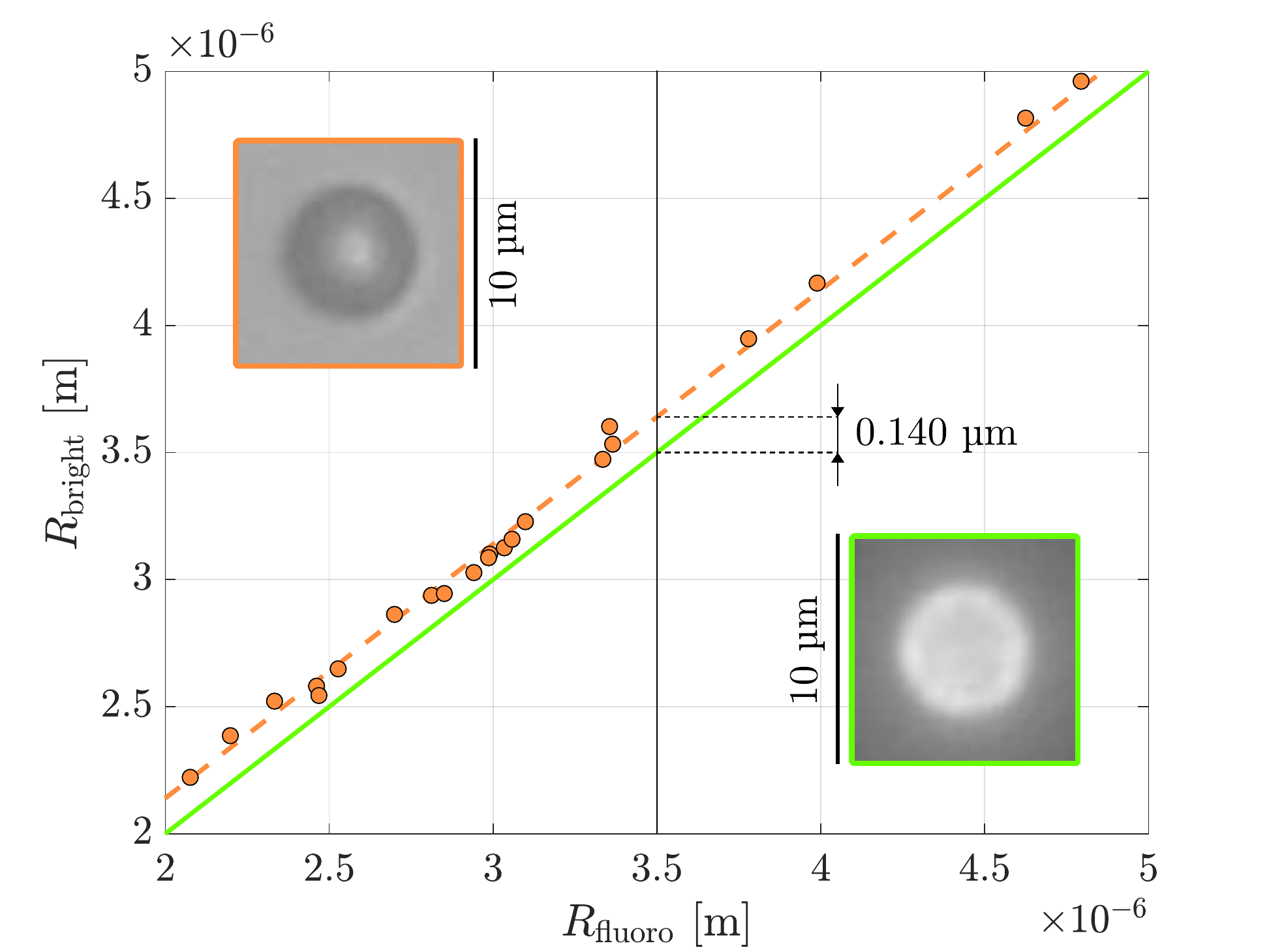}
    \caption{Comparison between bubble radii measured from a fluorescence ($R_{\rm fluoro}$) and bright-field ($R_{\rm bright}$) microscopy image (orange dots). The green solid line represents the zero-offset line, while the orange dashed line represents a fit with an offset of $\SI {140}{\nano\meter}$. Snapshots of the same microbubble taken with bright-field and fluorescence microscopy
 are shown in the orange and green boxes, respectively. }
    \label{fig:brightVfluoro}
\end{figure}

Measuring accurately the radius of a microbubble is essential to equitably compare its experimental behaviour with theoretical prediction.
However, making this measurement with bright-field microscopy is not trivial as the size of a microbubble is comparable to the wavelength of light.
The light waves passing around the microbubble generate a Fresnel diffraction pattern that appears as concentric bright and dark circles around the bubble and thus hinders the accurate estimation of its dimension.
Fortunately, fluorescence microscopy does not suffer from the interference between the light rays and the specimen because the specimen itself is self-emitting. 
By adding a fluorophore to the shell composition, it is possible to leverage fluorescence microscopy to image the microbubble.
However, the brightness intensity of fluorophores is not sufficient for ultra-high speed recordings.
Fluorescence microscopy is therefore only employed to evaluate the systematic error present in the bubble radius measurements obtained with bright-field microscopy. The error is assessed by comparing the bubble radius as measured with fluorescence and bright-field microscopy. Figure \ref{fig:brightVfluoro} compares the two, showing that bright-field microscopy overestimates the bubble radius by $R_{\rm err} = \SI[separate-uncertainty = true]{140(40)} {\nano\meter}$ across all bubble sizes.
It is important to mention that this error value is characteristic to this specific optical setup and follows from the particular combination of the light source and objective lens numerical apertures used.
It should not be surprising that a discrepancy smaller than the pixel resolution can be recovered, owing to the sub-pixel accuracy of the edge detection algorithm employed for detecting the bubble contour.

\subsection{Microbubble response upon ultrasonic driving}
\begin{figure*}[ht] 
    \centering
        \includegraphics[width=\textwidth]{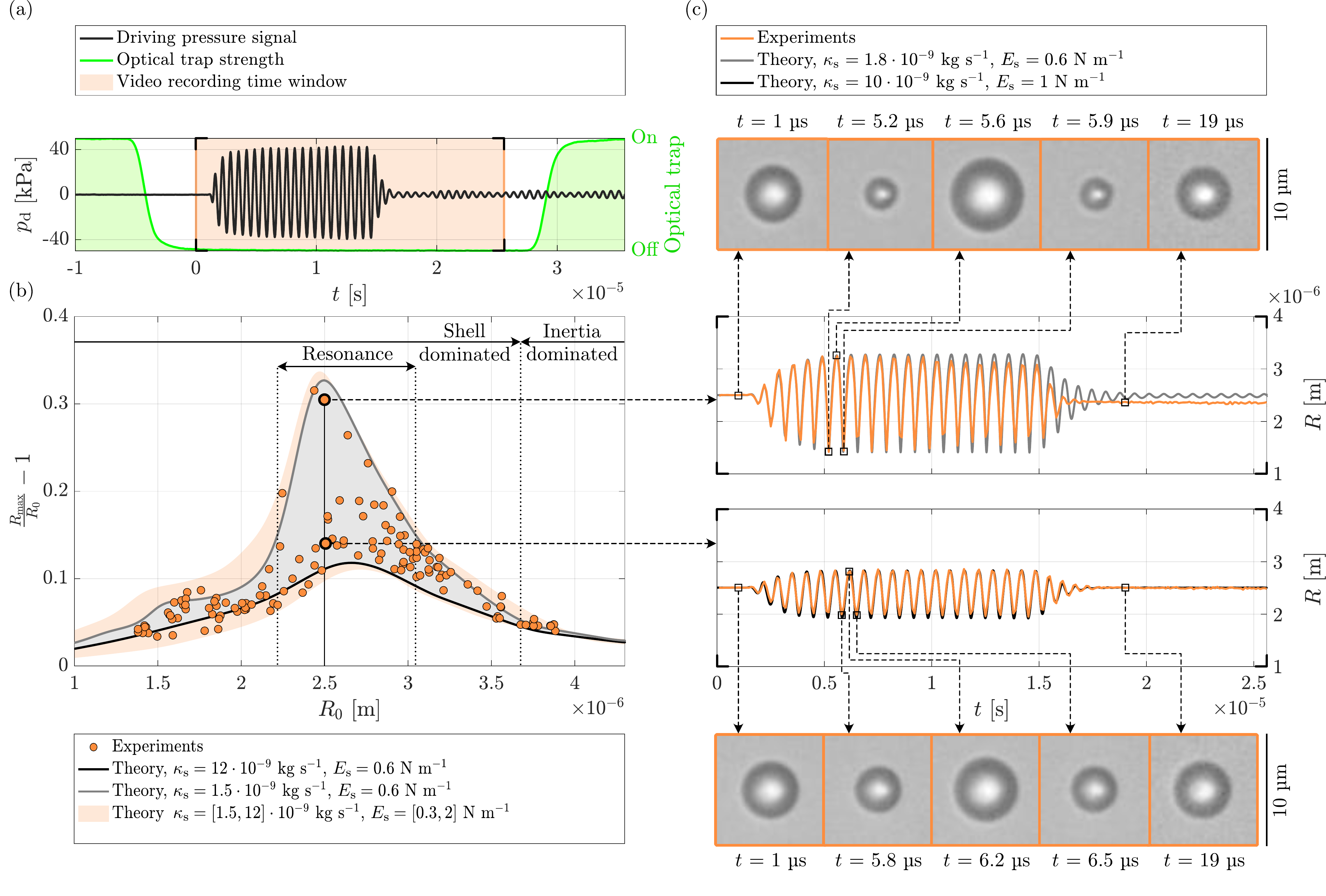}
\caption{(a) Sequence and time of events in the experiment.
The black line represents the driving pressure pulse produced by the ultrasound transducer and recorded with a hydrophone ($\SI{1.5}{\mega\hertz}$, $\SI{40} {\kilo\pascal}$, 20 cycles).
The green line shows the optical trap intensity around the microbubble in arbitrary units measured with a photodetector.
The orange-shaded area depicts the video recording time window.
(b)~Normalised maximum radial expansion for different equilibrium bubble radii $R_0$ extracted from time-resolved video recordings of single bubble responses to the driving pressure pulse (orange dots).
The orange-shaded area represents the envelope of maximum radial expansion simulated numerically with the model detailed in \S \ref{BubbleModel} using the range of shell modulus and viscosity values given in the legend.
The grey and black solid lines depict the numerically computed maximum radial expansion for a low and high shell viscosity, respectively, and a fixed shell modulus (values given in the legend).
(c) Radial time evolution of two bubbles having approximately the same size but largely different radial excursions (orange line).
Selected snapshots of the dynamics of the two bubbles, extracted from Supplementary Movies 1 and 2, are shown in the orange boxes.
The grey and black solid lines represent the numerically computed bubble radial time evolution for a low and high shell viscosity, respectively, and a fixed shell modulus (values given in the legend). 
The remaining parameters used in the simulations are: $\ p_0 = \SI{102.2} {\kilo\pascal}, \ \sigma_w = \SI{72.8} {\milli\newton\per\meter}, \ c_l = \SI{1481} {\meter\per\second}, \ \mu_l = \SI{9.54e-4} {\pascal\second}, \ \rho_l = \SI{997.8} {\kilo\gram\per\cubic\meter}, \sigma_0 = \SI{0} {\newton\per\meter}, \ \gamma = 1.4$, \ $K_g = \SI{0.026} {\watt\per\meter\per\kelvin}, \ \mathcal{R} = \SI{287} {\joule\per\kilo\gram\per\kelvin}, \ T_g|_R = \SI{295} {\kelvin}$.} 
    \label{fig:R0VE_RT}
\end{figure*}

The lipid-coated microbubbles are injected at a very low dilution ratio in the test chamber in order to avoid inter-bubble radiative forces (secondary Bjerknes forces) upon acoustic excitation.
Because of buoyancy, the bubbles float up and adhere to the top window of the test chamber.
Individual bubbles are then optically trapped and moved away from the top window by $ > \SI{50} {\micro\meter}$ using optical tweezers in order to avoid wall interferences.
The order of the optical vortex and the laser intensity are adjusted manually for each bubble size. 
The bubble under investigation is acoustically driven by a 20-cycle sinusoidal pulse at $ \SI{1.5} {\mega\hertz}$ and with $ \SI{40} {\kilo\pascal}$ of peak negative pressure produced by the ultrasound transducer.
A video recording at 10 million frames per second of the bubble response is performed using the ultra-high-speed videomicroscopy apparatus.
The optical trap is deactivated during the recording by using the microsecond-fast electronic shutter.
Figure \ref{fig:R0VE_RT}(a) summarises the temporal chain of events occurring during a single experiment, including the measured pressure driving pulse (black), the measured laser intensity (i.e.,\ the optical trap strength) (green) and the recording time window (orange). 

Radius-time curves are extracted from the recordings using a feature-extraction image-processing algorithm.
The normalised maximum bubble radial expansion ${R_{\rm max}}/{R_0} - 1$ is taken as the exemplary feature of the bubble response and as the target for the inverse problem of inferring the shell viscosity from the bubble response.
This choice is more robust than considering the average maximum expansion over several oscillation cycles because it prevents the bubble dissolution, which may occur in the final cycles to a varying extent, or shape modes, which develop for large bubbles again in the final cycles, from influencing the bubble response evaluation and the consequent shell viscosity estimation. 
In fact, bubble shrinkage leads to a reduction in radial excursion while shape modes cause the loss of the bubble's spherical symmetry and thus the inability to correctly relate the bubble's cross sectional area with its volume and hence to an equivalent radius.
Nevertheless, these issues always occur after the maximum bubble expansion is reached, which we can easily verify visually from individual videos.
By selecting the maximum bubble expansion as the target for the inverse problem of inferring the shell viscosity from the bubble response, we effectively prevent these concerns from compromising the estimation of shell viscosity.
Figure \ref{fig:R0VE_RT}(b) depicts with orange dots the distribution of the normalised maximum bubble radial expansion across a range of bubble equilibrium radii $R_0$.
The bubble radius is adjusted by considering the systematic sizing error coming from the Fresnel interference and deemed equal to $R_{\rm err} = \SI{140} {\nano\meter}$. 

As expected, the curve exhibits a pronounced peak at a specific bubble radius that represents the \emph{resonance bubble radius}.
However, it also unveils a large variability in the response for microbubbles of similar size.
A representative example of this arbitrariness is detailed in Fig.\ \ref{fig:R0VE_RT}(c), which shows the large difference in the radial time evolution of two bubbles of nearly identical size ($R_0 = \SI{2.5} {\micro\meter}$), reporting their radius-time curves in solid orange lines and selected snapshots of their dynamics, extracted from Supplementary Movies 1 and 2, in the orange boxes. 
This irregular bubble behaviour can be explained with a dispersion in the shell parameters: dilatational modulus but most prominently dilatational viscosity.
The orange-shaded area in Fig.\ \ref{fig:R0VE_RT}(b), computed using the bubble dynamics model reported in \S \ref{BubbleModel}, shows how a range of dilatational modulus $E_s = \SIrange{0.3}{2} {\newton\per\meter}$ and dilatational viscosity $\kappa_s = \SIrange {1.5e-9}{12e-9} {\kilo\gram\per\second}$ allows us to fully describe the variability in the experimental results.
This figure also suggests a useful qualitative classification of the bubble response into two main regimes, i.e.,\ \emph{shell-dominated} and \emph{inertia-dominated}, based on the bubble size.

In the \emph{shell-dominated} regime, the shell properties have a large influence on the bubble dynamics.
Therefore, minor changes in the shell rheology may lead to profoundly different bubble response outcomes.
This leverage effect is amplified in the \emph{resonance} sub-regime where a bubble experiences the largest radial excursion.
It should be noted that high radial excursion values can possibly result in the elastic shell regime to play a marginal role in the overall bubble response compared to the prevailing buckling and rupture regimes.
This is also the case for the combination of shell composition and negative peak pressure employed in this study.
In fact, as shown in Fig.\ \ref{fig:R0VE_RT}(b), the bubble response in the resonance regime is not influenced by the choice of the dilatational modulus.
The grey-shaded area, representing the dispersion of the bubble response for the same range in shell viscosity reported before but now for a specific dilatational modulus value $E_s = \SI{0.6}{\newton\per\meter}$, matches finely the orange-shaded one.
Therefore, for clinical applications, where the bubble suspension is subjected to relatively strong ultrasound forcing and the resonance is leveraged to get the maximum result at the minimum mechanical index, characterising the shell viscosity is certainly more important than doing so for the dilatational modulus.
Nonetheless, for sub-resonant bubbles, the response is less sensitive to shell viscosity variations and therefore these alone --- even with a larger range of values than the one used for Fig.\ \ref{fig:R0VE_RT}(b) --- cannot fully explain the scatter found in the experimental results, which must therefore also be attributed to the dilatational modulus.

On the other hand, in the \emph{inertia-dominated} regime, the shell plays a marginal role and the radial excursion of the bubble is mainly dependent on its equilibrium radius, and therefore on its inertia, as evident from the different curves representing different values of shell properties collapsing on a single one. 
As a consequence, the elevation of this curve can only be controlled by the driving pressure.
Therefore, finding a good match between experiments and theory in this regime means that the measured and theoretical pressure are coinciding, propping our acoustic pressure measurement.
Also note how the variability of the bubble response drops in this regime which correlates with the diminished influence of shell properties on the response itself.
This observation further reinforces the idea that the observed variability in the bubble response stems from the scattering of shell properties from one bubble to another.

The measured bubble radial time evolutions agree well with the theoretical predictions as shown by the representative radius-time curves in Fig.\ \ref{fig:R0VE_RT}(c).
The bubble with the largest excursion is best modelled with a shell viscosity $\kappa_s = \SI {1.8e-9} {\kilo\gram\per\second}$.
The discrepancy in the final cycles and the new equilibrium radius are caused by the net gas loss of the bubble, which is likely facilitated by the partial shedding of the shell.
These two phenomena are not accounted for in the theoretical model.
Consequently, as the ultrasound burst progresses, the bubble experiences changes in its resonance frequency and shell properties compared to the initial conditions, leading to deviations in its response from the theoretical predictions. 
Conversely, the bubble with the smallest radial excursion is simulated with a much higher shell viscosity $\kappa_s = \SI {10e-9} {\kilo\gram\per\second}$.
In this case, the bubble does not undergo shrinkage, resulting in excellent agreement between the theoretical and experimental curves, even in the tail region.
This result further strengthens our hypothesis regarding shell viscosity being the prime actor in explaining the variability in bubble response.
We emphasise that bubble dissolution only occurs during intense radial oscillations.
Otherwise, the bubbles remain stable for extended periods, tens of minutes, within the aqueous solution.

The theoretical bubble response is calculated by setting the initial surface tension $\sigma_0$ to zero because no finite value can encompass all data points, whereas a null value successfully does so, as illustrated in Fig.\ \ref{fig:R0VE_RT}(b).
It is worth noting that a tensionless or nearly tensionless (null Laplace pressure) bubble is also consistent with its long-term stability against dissolution in a saturated medium \cite{Ferrara2007UltrasoundDelivery}.
Although the microbubbles are originally fabricated with a perfluorobutane gas core, it is likely that this has been replaced by air due to the extended period of time (few tens of minutes) that the bubbles reside in an air-saturated aqueous medium.
A study by Kwan and Borden \cite{Kwan2012LipidExchange}, supports this notion, revealing that the heavy-molecular-weight gas initially present in the gas core of microbubbles is substituted by the air dissolved in the surrounding aqueous medium within a few minutes.
Accordingly, we set $\gamma = 1.4$, $K_g = \SI{0.026} {\watt\per\meter\per\kelvin}$ and $\mathcal{R} = \SI{287} {\joule\per\kilo\gram\per\kelvin}$.
The driving acoustic signal used in the simulations is the actual experimentally recorded pressure pulse [Fig.\ \ref{fig:R0VE_RT}(a)].
The remaining parameters used are: $p_0 = \SI{102.2} {\kilo\pascal}, \ \sigma_w = \SI{72.8} {\milli\newton\per\meter}, \ c_l = \SI{1481} {\meter\per\second}, \ \mu_l = \SI{9.54e-4} {\pascal\second}, \ \rho_l = \SI{997.8} {\kilo\gram\per\cubic\meter}, \ T_g|_R = \SI{295} {\kelvin}$.

\subsection{Shell dilatational viscosity estimation}
The shell dilatational viscosity of each tested microbubble is inferred from the observation of its radial time evolution by solving an inverse problem.
A grid search optimisation algorithm is used to find the theoretical bubble radial time evolution that results in the minimum deviation between empirical and theoretical maximum bubble expansions.
The free parameters are the dilatational shell viscosity $\kappa_s$ and modulus $E_s$.
The other parameters are held fixed at the values presented in the previous section.
We would like to emphasise that due to the typical shape of pressure signals generated by ultrasound transducers, inferring shell viscosity from the maximum expansion of the bubble is a more robust approach than inferring it from the decaying rate of the bubble response tail.
The response tail, in fact, does not truly represent the free dissipative relaxation of the bubble but rather the forced response to the tail of the pressure driving signal [see Fig. ~\ref{fig:R0VE_RT}(a)].
On the contrary, the (maximum) radial expansion of a bubble correlates robustly with its shell viscosity as shown in Fig. ~\ref{fig:R0VE_RT}(b).

\begin{figure}[ht] 
    \centering
        \includegraphics[width=\columnwidth]{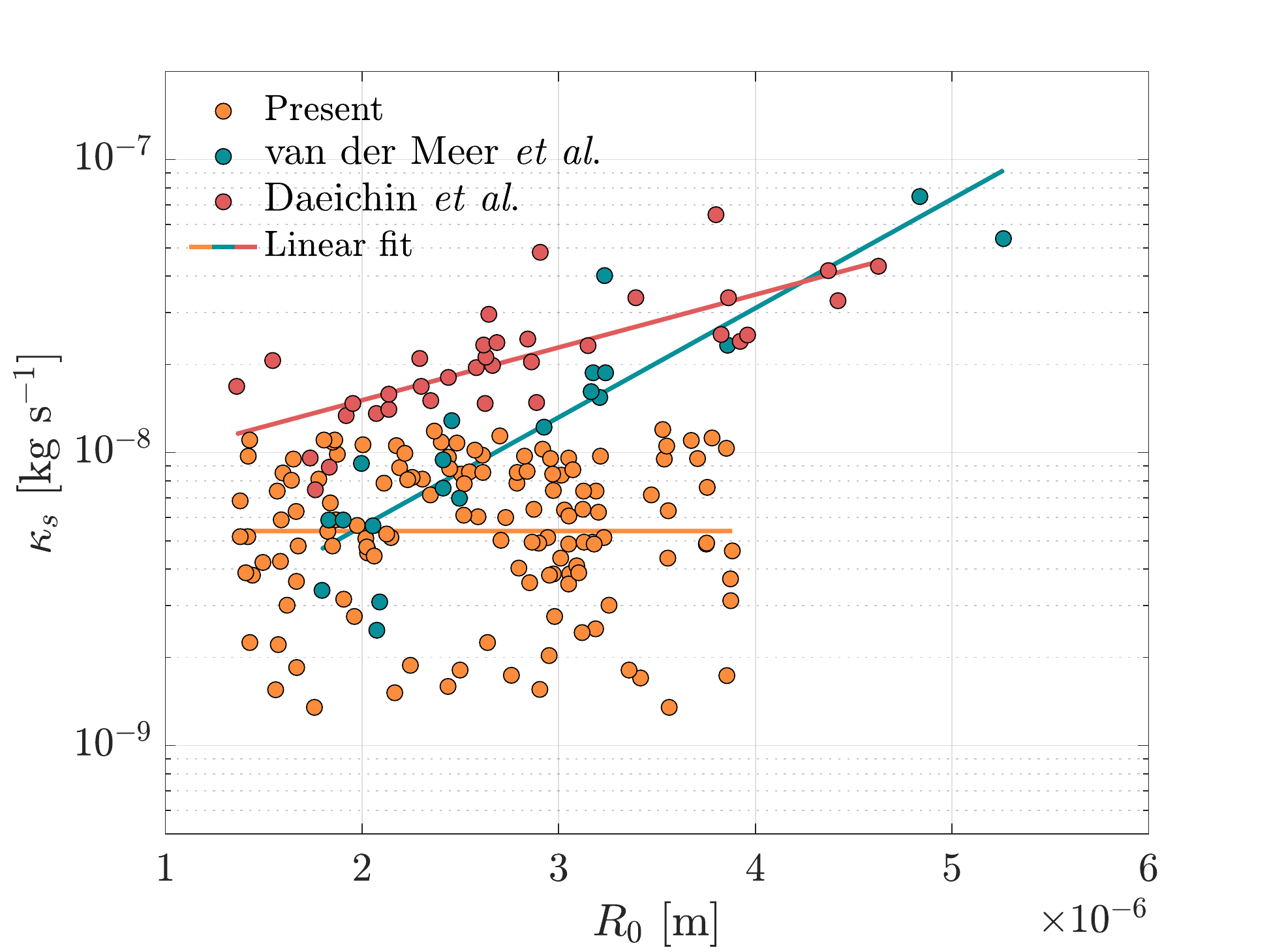}
   \caption{Shell dilatational viscosity $\kappa_s$ for different equilibrium bubble radii $R_0$.
    The orange dots represent the values gathered in the present study by solving the inverse problem with an optimisation algorithm. 
    The objective function is to minimise the deviation in the maximum normalised radial expansion between the experimental and synthetic radial time evolutions.
    The green dots represent the dataset collected by van der Meer \textit{et al}.     \cite{VanDerMeer2007MicrobubbleAgents} by extracting the shell viscosity from the width of the frequency response of single microbubbles reconstructed by scanning through a range of driving frequencies.
    The red dots represent the dataset collected in Daeichin \textit{et al}. \cite{Daeichin2021PhotoacousticAgents} by extracting the shell viscosity from the width of the frequency response of single microbubbles directly acquired using impulsive driving.
    The coloured solid lines represent linear fits to the corresponding datasets.}
    \label{fig:R0Ks}
\end{figure}

Figure~\ref{fig:R0Ks} reports with the orange dots the values of shell viscosity resulting from solving the inverse problem, which show a large variability of one order of magnitude and an average value of $\overline{ \kappa_s} =\SI{5.4e-9} {\kilo\gram\per\second}$.
In contrast to previous studies (e.g. in red van der Meer \textit{et al}. \cite{VanDerMeer2007MicrobubbleAgents}, in green Daeichin \textit{et al}. \cite{Daeichin2021PhotoacousticAgents}), no dependence on the bubble size is found.

We feel confident about the reliability of our shell viscosity estimates, i.e.\ the absence of biases potentially being introduced by the theoretical model used to solve the inverse problem, because we have proven with Fig.\ \ref{fig:R0VE_RT}(c) that the model employed is capable of describing the (nonlinear) dynamics of \emph{single} microbubbles and, most importantly, with Fig.\ \ref{fig:R0VE_RT}(b), that it is also able to faithfully represent the \emph{overall} microbubble behaviour across the whole size range with a single set of parameters, and in particular, to match the position of the experimental resonance peak, which is crucial for avoiding systematic errors.
From this perspective, one can finally understand the importance of taking into account even the small systematic error $R_{\rm err}$ in the sizing of the bubbles, which causes a minimal but still significant shift in the resonance peak position.

It should be noted that these values of shell viscosity do not necessarily represent the ground truth, but instead, a single-number portrayal of the complex dissipative processes occurring in the shell through the lens of the Marmottant model.
Nothing prevents new rheological developments concerning the effect of dynamical parameters (e.g.\ the strain of the surface area) on the shell viscosity, from changing this quantitative picture.

We do not provide the result of the inverse problem regarding the dilatational modulus because it would be inconclusive.
As explained in the previous section, at these acoustic driving conditions, for a wide range of bubble sizes, the elastic regime of the shell plays a marginal role and therefore the modulus can take on an arbitrary value.
To correctly infer the modulus, it is necessary to probe the linear response of the bubble.
Unfortunately, resolving such small oscillations exceeds the limits of optical microscopy.
One must therefore resort to laser scattering techniques, which, however, compromise on the side of versatility.
Notwithstanding, the range of values found in this study, as shown in Fig.\ \ref{fig:R0VE_RT}(b), matches well the measurements carried out with the aforementioned laser scattering techniques (see e.g. Daeichin \textit{et al}. \cite{Daeichin2021PhotoacousticAgents}).

\section{Discussion}

\subsection{The shell dilatational viscosity shows large variability and no dependency on the bubble size}
We believe that the variability in the bubble response is genuine and not a byproduct of the experimental methodology in light of  the following arguments: (i) absence of significant arbitrary interference on the bubble dynamics from the pressure waves that reflect off from the top surface of the test chamber owing to the comparable acoustic impedances of water and the membrane material (low-density polyethylene, $z=\SI{1.79e6}{\pascal\second\per\meter}$) and as confirmed by the pressure pulse recorded inside the test chamber with a needle hydrophone inserted through a side opening and shown in Fig.\ \ref{fig:R0VE_RT}(a); (ii) absence of  significant arbitrary interference on the bubble dynamics from pressure waves being scattered by neighbouring bubbles by reason of the extremely low dilution ratio; (iii) absence of arbitrary thermal effects affecting the shell rheological properties caused by the optical trap  because of the very low localised temperature increase, estimated in the order of a few tens of millikelvin based on the work by Peterman \textit{et al.} \cite{Peterman2003Laser-inducedTraps} and the very low electromagnetic absorption coefficient of water at the wavelength of $\SI{532} {\nano\meter}$ \cite{Hale1973OpticalRegion,Pope1997AbsorptionMeasurements}.

Most probably, the variability in bubble response can be rooted in the arbitrariness of the rheological properties of the shell. 
In support of this theory, it is noteworthy that the variability of the bubble response changes in accordance with the influence exerted by the shell properties on it, as clearly illustrated in Fig.\ \ref{fig:R0VE_RT}(b).
The rheological properties of the shell, in turn, are dependent on its microstructure.
At surface tension values close to zero, a phospholipid monolayer self-assembles in a two-dimensional polydomain liquid crystal with different tailgroup orientations.
High variation in the size and distribution of such phospholipid domains on the bubble interface has been observed by Borden \textit{et al.} \cite{Borden2006LateralMicrobubbles} by using fluorescence and electron microscopy and by Kooiman \textit{et al.} \cite{Kooiman2010LipidMicrobubbles} with confocal fluorescence microscopy.
The edge of the domains, where phospholipids with different tailgroups interact with each other, can be considered as defects of the bubble shell.
A larger amount of defects results in more frequent interactions between phospholipids with different tailgroup orientations which, in turn, could yield a higher shell viscosity as experimentally proven by Hermans \& Vermant \cite{Hermans2014InterfacialConditions} by testing the interfacial shear rheology of DPPC at different preshear values (i.e.\ at different levels of crystalline microstructure refinement).
Based on these findings, we can conclude that the rheological properties of the shell and consequently the acoustic behaviour of the bubble are uniquely determined by the characteristic processing history of the shell.
It follows that we cannot rule out the possibility that the repeated driving pressure pulses to which the bubbles in the test chamber are subjected during the test may change the original morphology of the shell (through clustering, partitioning or even shedding off of shell material). 
Nevertheless, the time interval between each ultrasound pulse in our experiment (a few minutes) is designed to allow for the completion of any dissolution processes and the restoration of full lipid coverage at the bubble interface (i.e.\ null surface stresses).
Therefore, we believe that the effect of shell aging, which inevitably occurs during the operation of microbubbles in any case, does not significantly contribute to the observed variability in shell properties that is already naturally present.

The independence of shell viscosity on the bubble size found in our measurements is consistent with the unchanging shell thickness (consisting, most probably, of only a monolayer of phospholipids) and therefore shell properties with the size of the bubble. 
However, our shell viscosity estimates contradict previous studies that reported a smaller scatter and an upward trend in shell viscosity as the bubble equilibrium radius increases, albeit with varying rates across studies\cite{Morgan2000ExperimentalSize,VanDerMeer2007MicrobubbleAgents,Doinikov2009ModelingMicrobubbles,Tu2009EstimatingScattering,Tu2011MicrobubbleCytometry,Helfield2013NonlinearSpectroscopy, Parrales2014AcousticProperties,vanRooij2015Non-linearDPPC,Segers2018High-precisionMicrobubbles,Supponen2020ThePopulations,Daeichin2021PhotoacousticAgents}. 
We believe that these divergent results are an artifact caused by the measurement techniques adopted.
In the next sections, we present arguments supporting our hypothesis.

\subsection{Bubble spectroscopy: overestimating the frequency response bandwidth  affects the shell viscosity estimation}\label{SpectroscopyProblems}
Most of studies which found the upward trend with the bubble equilibrium radius have estimated the shell viscosity by analysing the bubble response in the frequency domain \cite{VanDerMeer2007MicrobubbleAgents,Helfield2013NonlinearSpectroscopy, Parrales2014AcousticProperties,vanRooij2015Non-linearDPPC,Segers2018High-precisionMicrobubbles,Supponen2020ThePopulations,Daeichin2021PhotoacousticAgents}, adopting the "bubble spectroscopy" technique introduced by van der Meer \textit{et al}. \cite{VanDerMeer2007MicrobubbleAgents} or an equivalent of it for acoustic attenuation measurements \cite{Parrales2014AcousticProperties,Segers2018High-precisionMicrobubbles}.
The bubble spectroscopy method propounds to estimate the shell viscosity from the half-energy bandwidth of the microbubble frequency response treating the microbubble as a linear harmonic oscillator.
We now show how the upward trends are caused by the overestimation of bandwidth due to experimental biases.
To align ourselves with these previous works and for ease of exposition from here on, we adopt the polytropic process approximation for the gas contained in the bubble [i.e.,\ Eq.\ \eqref{eq:PolyGas}].

If one assumes that the bubble undergoes small radial oscillations, Eq.\ \eqref{eq:RP} can be linearised and reduced to:
\begin{equation} \label{eq:LinearOscillator}
\ddot x +2\delta\omega_0\dot x + \omega_0^2 x = -p_d(t)/\rho_l R_0^2, 
\end{equation}
where $x(t) = \left(R(t)-R_0\right)/R_0$ is the normalised radial excursion, $\omega_{0}=2\pi f_0$ is the natural angular frequency of the system expressed as
\begin{equation}
\omega_0 = \left[\frac{1}{\rho_l R_0^2}\left(3n p_{\infty} + \frac{2 (3n-1) \sigma_0}{R_0} + \frac{4E_s}{R_0}\right)\right]^{1/2},
\end{equation}
and $\delta$ is the damping coefficient consisting of three contributions: the contribution from sound reradiation
\begin{equation}
\delta_{\rm rad} = \frac{3n p_{\infty} R_0 + 6\sigma_0 n}{2\rho_l R_0^2 \omega_0 c_l},
\end{equation}
the contribution from the liquid viscosity
\begin{equation}
 \delta_{\rm vis} = \frac{2\mu_l}{\rho_l R_0^2 \omega_0},
\end{equation}
and the contribution from the shell dilatational viscosity
\begin{equation}
\delta_{\rm shell} = \frac{2\kappa_s}{\rho_l R_0^3 \omega_0}.
\end{equation}
Assuming a harmonic acoustic driving $p_d(t) = p_a\sin(\omega t)$ and applying the Fourier transform to Eq.\ \eqref{eq:LinearOscillator}, one obtains the frequency response $\mathcal{H}(\omega)$ of the bubble-oscillator:
\begin{equation} \label{eq:LinearOscillatorFourier}
\mathcal{H}(\omega) = \frac{x(\omega)}{p_d (\omega)} = - \frac{ (\rho_l R_0^2)^{-1} }{-\omega^2  +2i\delta\omega_0 \omega   + \omega_0^2  },
\end{equation}
where $\omega = 2 \pi f$ is the ultrasound angular frequency.
If the damping $\delta$ is small enough ($\delta < 0.1$), this can be derived from the frequency response energy $\left|\mathcal{H}(\omega)\right|^2$ as
\begin{equation}\label{eq:SpectroscopyDamping}
\delta \approx \frac{\rm FWHM_{\mathcal{H}^2}}{2f_{\rm res}}, 
\end{equation}
where $\rm FWHM_{\mathcal{H}^2}$ is the full width of the frequency response energy at half peak response (half-energy bandwidth) and $f_{\rm res} = f_0(1-2\delta^2)^{1/2}$ is the peak response frequency or \textit{resonance frequency}.
Finally, the shell viscosity $\kappa_s$ can be derived from the damping $\delta$ as
\begin{equation}\label{eq:Spectroscopyks}
\kappa_s = \frac{1}{2}\rho_l R_0^3\omega_0(\delta-\delta_{\rm rad}-\delta_{\rm vis}). 
\end{equation}

\begin{figure}[!ht] 
    \centering
        \includegraphics[width=\columnwidth]{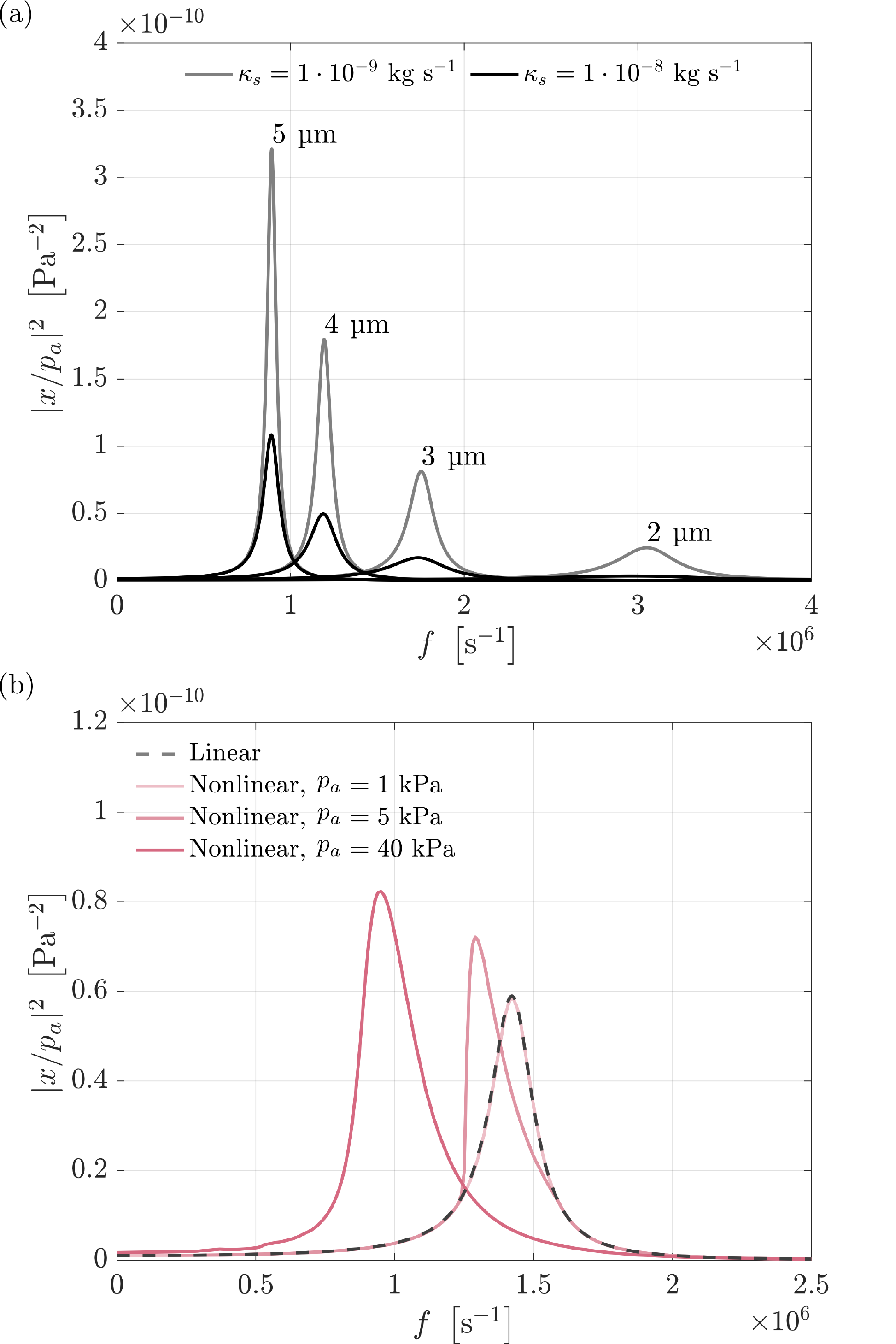}
    \caption{(a) Frequency response energies computed with the linearised model [Eq.\ \eqref{eq:LinearOscillatorFourier}] for microbubbles with equilibrium radii $R_0 = \SIrange{2}{5} {\micro\meter}$ and shell viscosities $\kappa_s = \SI{1e-9}{\kilo\gram\per\second}$ (grey line) and $\SI{1e-8} {\kilo\gram\per\second}$ (black line).
    (b) Effect of the acoustic driving pressure amplitude $p_a$ on the frequency response energy of a bubble with equilibrium radius $R_0 = \SI{3.5}{\micro\meter}$ and shell viscosity $\kappa_s = \SI{5e-9}{\kilo\gram\per\second}$.
    The grey dashed line is the result of the linearised model [Eq.\ \eqref{eq:LinearOscillatorFourier}].
    The pink solid lines are the results of the modified Rayleigh--Plesset equation [Eq.\ \eqref{eq:RP}] with the Marmottant shell model [Eq.\ \eqref{eq:InterfacePressureMarmottant}] and the polytropic process approximation for the gas contained in the bubble [Eq.\ \eqref{eq:PolyGas}] for increasing acoustic driving pressure amplitude $p_a = \SI{1} {\kilo\pascal}, \   \SI{5} {\kilo\pascal} \ \text{and} \ \SI{40} {\kilo\pascal}$. 
    The remaining physical parameters used, so as to align with earlier works such as van der Meer \textit{et al.} \cite{VanDerMeer2007MicrobubbleAgents}, are: $p_0 = \SI{100} {\kilo\pascal}, \ \sigma_w = \SI{72} {\milli\newton\per\meter}, \ c_l = \SI{1500} {\meter\per\second}, \ \mu_l = \SI{0.002} {\pascal\second}, \ \rho_l = \SI{1000} {\kilo\gram\per\cubic\meter}, \ E_s = \SI{0.54} {\newton\per\meter}, \ \sigma_0 = \SI{0.02} {\newton\per\meter},  \ n = 1.07$.}
    \label{fig:FrequencyResponsePlot}
\end{figure}

Exemplary frequency response energies $\left|\mathcal{H}(\omega)\right|^2$ for microbubbles obeying the linearised model [Eq.\ \eqref{eq:LinearOscillatorFourier}], with equilibrium radii $R_0 = \SIrange{2}{5} {\micro\meter}$ and shell viscosities $\kappa_s = \SI{1e-9}{\kilo\gram\per\second}$ and $\SI{1e-8} {\kilo\gram\per\second}$, are given in Fig.\ \ref{fig:FrequencyResponsePlot}(a).
These two values are chosen because they represent approximately the limit values of shell viscosity found in the present study. 
It is good to be reminded, however, that this linear approximation only holds for markedly small acoustic driving.
Based on the Marmottant shell model, as early as $p_a = \SI{5} {\kilo\pascal}$, the nonlinear frequency response deviates from the linear one, as shown in Fig.\ \ref{fig:FrequencyResponsePlot}(b) for a microbubble with radius $R_0 = \SI{3.5} {\micro\meter}$ and shell viscosity $\kappa_s = \SI{5e-9}{\kilo\gram\per\second}$.
Furthermore, this threshold in $p_a$ decreases the closer the microbubble at rest is to the buckled state (i.e.,\ in Marmottant model terms, the closer $\sigma_0$ is to zero).
This suggests that the statement made in the study of van der Meer \textit{et al.} \cite{VanDerMeer2007MicrobubbleAgents} regarding the absence of substantial differences in the frequency response of a microbubble driven at $p_a = \SI{1} {\kilo\pascal}$ and one at $p_a = \SI{40} {\kilo\pascal}$ is inaccurate.
This oversight was caused by the use of a linearised model to simulate the frequency response of the microbubble driven at $p_a = \SI{40} {\kilo\pascal}$.
%
% Figure obtained from "PowerTransferFunctionPlot.m" script and "ComparisonLinearVNonlinearPlot.m" script

\begin{figure}[!ht] 
    \centering
        \includegraphics[width=\columnwidth]{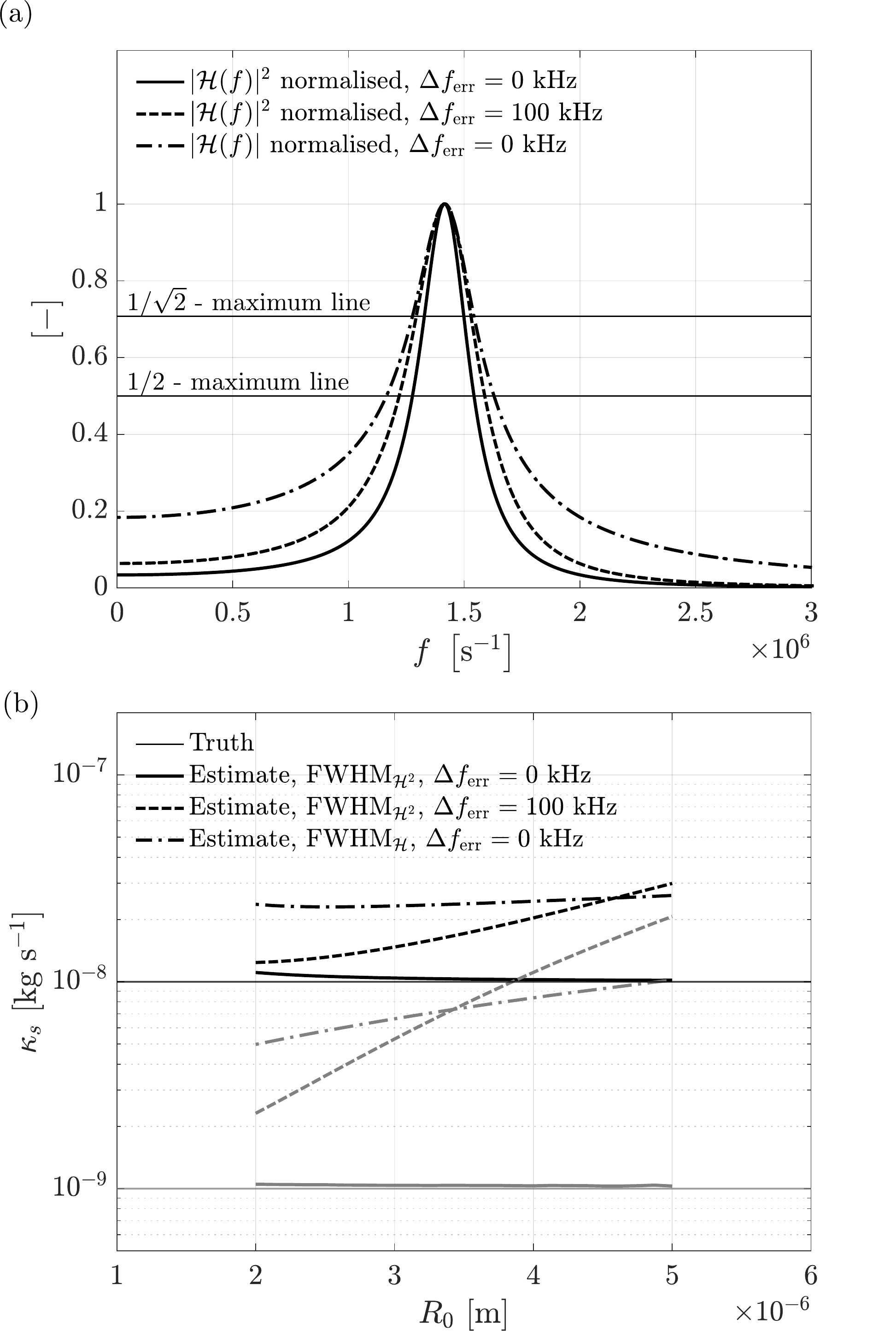}
    \caption{Effectiveness of the bubble spectroscopy method in inferring the shell viscosity of microbubbles with frequency responses affected by errors.
    (a) The solid line represents the error-free frequency response energy $\left|\mathcal{H}(\omega)\right|^2$ of a microbubble that obeys the linear model [Eq.\ \eqref{eq:LinearOscillatorFourier}].
    The dashed line represents the frequency response energy $\left|\mathcal{H}(\omega)\right|^2$ of the same bubble but affected by a bandwidth overestimation $\Delta f_{\rm err} = \SI{100}{\kilo\hertz}$.
    The dotted-dashed line represents the error-free frequency response magnitude $\left|\mathcal{H}(\omega)\right|$ of the same bubble.
    All the lines pertain to a bubble with equilibrium radius $R_0 = \SI{3.5} {\micro\meter}$ and shell viscosity $\kappa_s = \SI{1e-8}{\kilo\gram\per\second}$.
    (b) The thin solid lines represent the shell viscosity values to be inferred, in black $\kappa_s = \SI{1e-8}{\kilo\gram\per\second}$ and in grey $\kappa_s = \SI{1e-9}{\kilo\gram\per\second}$.
    The thick solid lines are the result of applying the bubble spectroscopy method [Eqs. \eqref{eq:SpectroscopyDamping} and \eqref{eq:Spectroscopyks}] to error-free $\left|\mathcal{H}(\omega)\right|^2$.
    The dashed lines depict the result of bubble spectroscopy applied to $\left|\mathcal{H}(\omega)\right|^2$ affected by $\Delta f_{\rm err} = \SI{100}{\kilo\hertz}$.
    The dotted-dashed lines depict the result of bubble spectroscopy applied to $\left|\mathcal{H}(\omega)\right|$ instead of $\left|\mathcal{H}(\omega)\right|^2$.
    The physical parameters used are: $p_0 = \SI{100} {\kilo\pascal}, \ \sigma_w = \SI{72} {\milli\newton\per\meter}, \ c_l = \SI{1500} {\meter\per\second}, \ \mu_l = \SI{0.002} {\pascal\second}, \ \rho_l = \SI{1000} {\kilo\gram\per\cubic\meter}, \ E_s = \SI{0.54} {\newton\per\meter}, \ \sigma_0 = \SI{0.02} {\newton\per\meter},  \ n = 1.07$.}
    \label{fig:BandwidthErrors}
\end{figure}

In an experimental investigation, the frequency response of a microbubble can be reconstructed by scanning through a range of driving frequencies \cite{VanDerMeer2007MicrobubbleAgents,Luan2012AcousticalMicrobubbles,Helfield2013NonlinearSpectroscopy,vanRooij2015Non-linearDPPC,Lum2018PhotoacousticProperties,Supponen2020ThePopulations} or directly acquired using impulsive driving \cite{Daeichin2021PhotoacousticAgents}.
If the response thus obtained follows the linearised model described by Eq.\ \eqref{eq:LinearOscillatorFourier} and depicted in Fig.\ \ref{fig:BandwidthErrors}(a) in the solid line, the physical parameters are known exactly and the response is free of experimental biases, then the bubble spectroscopy method, through the use of Eqs. \eqref{eq:SpectroscopyDamping} and \eqref{eq:Spectroscopyks}, allows the shell viscosity to be inferred correctly, as shown for $\kappa_s = \SI{1e-9}{\kilo\gram\per\second}$ and $\SI{1e-8} {\kilo\gram\per\second}$ in Fig.\ \ref{fig:BandwidthErrors}(b) in solid grey and black lines, respectively.
However, if the experimental parameters used to survey the microbubble frequency response are not chosen carefully, the spectroscopy method can lead to an erroneous evaluation of the shell viscosity.
Large frequency steps when scanning through frequencies, short observation times of the bubble dynamics leading to spectral leakage and high driving pressure amplitudes that would negate the linear oscillator hypothesis can cause an overestimation of the half-energy bandwidth and, as a consequence, of the shell viscosity.
If we assume that these inaccuracies predominantly affect the half-energy bandwidth by a systematic deviation $\Delta f _{\rm err}$, the estimated shell viscosities will be impacted by an amount $\kappa_{s,{\rm err}} = \left(\pi/2\right) \rho_l R_0^3\Delta f _{\rm err}\left(1-2\delta^2\right)^{-1}$.
For small damping values, which is usually the case for phospholipid-coated microbubbles, this expression can be approximated as:
\begin{equation}\label{eq:viscosity_bandwidth_error}
    \kappa_{s,{\rm err}} \approx \frac{\pi}{2} \rho_l R_0^3\Delta f _{\rm err}.
\end{equation}
It is clear that the cubic dependency on the equilibrium bubble radius implies that even small systematic errors on the half-energy bandwidth yield significant variations in the shell viscosity across a range of bubble sizes.
The impact on the shell viscosity estimate of an experimental systematic error $\Delta f_{\rm err} = \SI{100}{\kilo\hertz}$ in the half-energy bandwidth of frequency responses, as illustrated in Fig.\ \ref{fig:BandwidthErrors}(a) in the dashed line,  is given in
Fig.\ \ref{fig:BandwidthErrors}(b) as dashed lines. 
This result can explain the unphysical increase of shell viscosity with the bubble size and the lower variability compared to the present work that many studies have observed.

We would also like to clarify, in view of some confusion in previous studies, that to correctly estimate the damping $\delta$ from the frequency response full width at half maximum (FWHW)  the frequency response \emph{energy} $\left|\mathcal{H}(\omega)\right|^2$ has to be used.
To estimate the damping from the frequency response \emph{magnitude} $\left|\mathcal{H}(\omega)\right|$, showcased in Fig.\ \ref{fig:BandwidthErrors}(a) in the dotted-dashed line, the full width at $1/\sqrt{2}$-maximum must be employed instead.
The consequences of estimating the damping from the half-magnitude bandwidth  instead of the half-energy bandwidth can be evaluated computing from Eq.\ \eqref{eq:LinearOscillatorFourier} the frequency at half magnitude $f_{\rm HM} = f_{\rm res}\left( \frac{1 \pm 2 \sqrt{3}\delta\sqrt{1-\delta^2}}{1-2\delta^2} \right)^{1/2}$, which for small values of damping can be approximated as $f_{\rm HM} \approx f_{\rm res}\left(1 \pm \sqrt{3} \delta \right)$, as opposed to the frequency at half energy $f_{\rm HE} \approx f_{\rm res}\left(1 \pm \delta \right)$. This translates into an error in the bandwidth $\Delta f_{\rm err} = 2\left(\sqrt{3}-1\right)f_{\rm res}\delta$ and a departure from the true value of the shell viscosity equal to:
\begin{equation}\label{eq:viscosity_magnitudespectrum_error}
    \kappa_{s,{\rm err}} = \frac{\sqrt{3} -1}{2} \rho_l R_0^3 \omega_0  \delta.
\end{equation}
The effect of this error on the shell viscosity estimate is given in Fig.\ \ref{fig:BandwidthErrors}(b) as dotted-dashed lines.
Again, the error causes an upward shift and an upward trend in the shell viscosity estimate as the equilibrium radius of the bubble increases.

% Figure obtained from "LinearRPTransferFunction.m" script

We also emphasise that an inappropriate choice of the bubble dynamics model or inaccuracies in the physical parameter measurements can also lead to unphysical trends in the shell viscosity.

We now examine two previous studies, in which the bubble spectroscopy method was employed to infer the shell viscosity, in order to verify whether the values found are correct or are, instead, possibly a byproduct of the inaccuracies listed in the previous section.
The two studies chosen are those of van der Meer \textit{et al.} \cite{VanDerMeer2007MicrobubbleAgents} and Daeichin \textit{et al.} \cite{Daeichin2021PhotoacousticAgents} because they stand as exemplary models for a large class of other studies which employed the bubble spectroscopy method. 
The values of shell viscosity they found are shown in Fig.\ \ref{fig:R0Ks}, along with the values from the present study.
The work of van der Meer \textit{et al.} pioneered the use of bubble spectroscopy and represents the linchpin for other closely related studies (e.g. Luan \textit{et al.} \cite{Luan2012AcousticalMicrobubbles}, Helfield \textit{et al.} \cite{Helfield2013NonlinearSpectroscopy} and van Rooij \textit{et al.} \cite{vanRooij2015Non-linearDPPC}).
The authors reconstructed the frequency response of single microbubbles point-by-point by scanning the frequency of the driving acoustic pulse produced by a broadband ultrasound transducer.
At each input frequency, the bubble dynamics was recorded by using an ultra-high-speed videomicroscopy facility, the radial time evolution extracted from it and its Fourier transform computed to finally obtain the response at the input frequency. 
The work of Daeichin \textit{et al.} represents the last and most developed iteration of the laser-scattering measurement technique, pioneered by Dove \text{et al.} \cite{ Dove2014OpticallyMicrobubbles} and already employed by Lum \textit{et al.}\cite{Lum2018PhotoacousticProperties} and Supponen \textit{et al.} \cite{Supponen2020ThePopulations}.
The authors here have developed a brilliant technique to characterise in one fell swoop the whole microbubble frequency response. The microbubble was probed with a broadband photoacoustic wave generated by a nanosecond-pulsed laser illuminating an optical absorber and the resulting oscillations were detected by light scattering of a continuous laser. 

\begin{figure}[!ht] 
    \centering
        \includegraphics[width=\columnwidth]{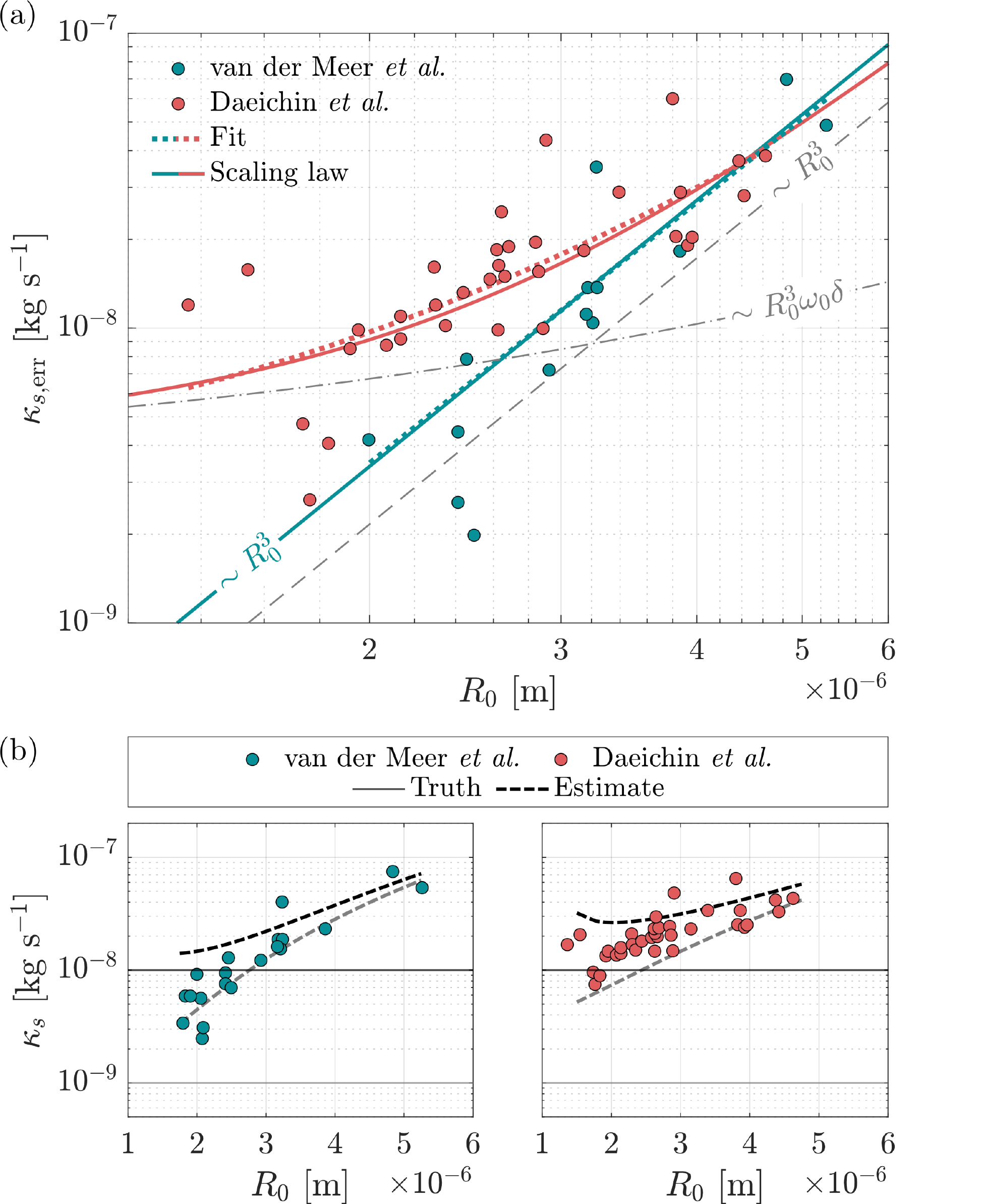}
    \caption{(a) Scaling laws of the deviation of the shell viscosity values reported in two previous studies from the assumed true value $\kappa_{s, \rm true} = \SI{5e-9} {\kilo\gram\per\second}$.
    The experimental values are given with the dots (van der Meer \textit{et al.} in green and Daeichin \textit{et al.} in red).
    The dotted lines represent a fit of the experimental values (linear in green, quadratic in red).
    The green solid line scales with $R_0^3$, representing the deviance caused by an error in the $\rm FWHM_{\mathcal{H}^2}$ [Eq.\ \eqref{eq:viscosity_bandwidth_error}].
    The red solid line results from the sum of a first contribution that scales with $R_0^3\omega_0\delta$ (dotted-dashed grey line) and a second one that scales with $R_0^3$ (dashed grey line). 
    The latter one represents the deviance caused by measuring the FWHM from the frequency response magnitude instead of energy [Eq.\ \eqref{eq:viscosity_magnitudespectrum_error}].
    (b) (left) Comparison between the shell viscosity values reported in van der Meer \textit{et al.} (green dots) and the values inferred from linear frequency response energies affected by an error $\Delta f_{\rm err} = \SI{270}{\kilo\hertz}$ in the $\rm FWHM_{\mathcal{H}^2}$ (dashed lines). The black lines correspond to a shell viscosity $\kappa_s = \SI{1e-8} {\kilo\gram\per\second}$, while the grey lines to a shell viscosity $\kappa_s = \SI{1e-9} {\kilo\gram\per\second}$.
    (right) Comparison between the shell viscosity values reported in Daeichin \textit{et al.} (red dots) and the values inferred from linear frequency response magnitudes affected by an error $\Delta f_{\rm err} = \SI{190}{\kilo\hertz}$ in the $\rm FWHM_{\mathcal{H}}$ (dashed lines).
    The physical parameters used are: $p_0 = \SI{100} {\kilo\pascal}, \ \sigma_w = \SI{72} {\milli\newton\per\meter}, \ c_l = \SI{1500} {\meter\per\second}, \ \mu_l = \SI{0.002} {\pascal\second}, \ \rho_l = \SI{1000} {\kilo\gram\per\cubic\meter}, \ E_s = \SI{0.54} {\newton\per\meter}, \ \sigma_0 = \SI{0.02} {\newton\per\meter},  \ n = 1.07$.}
    \label{fig:FigureKappasScalingLaws}
\end{figure}

As a first step, we study the scaling laws of the possible deviations of the shell viscosity values reported in the two studies from an assumed true value in order to identify the nature of the potential errors made. 
In view of our results reported in \S \ref{Results}, we assume a  value $\kappa_{s, \rm true} = \SI{5e-9} {\kilo\gram\per\second}$, constant across bubble sizes, as the true shell viscosity value.
Figure \ref{fig:FigureKappasScalingLaws}(a) reveals that the values found by van der Meer \textit{et al.} deviate from  $\kappa_{s, \rm true}$ following a trend (dotted green line) that varies proportionally to $R_0^3$ (solid green line).
This tendency, in view of Eq.\ \eqref{eq:viscosity_bandwidth_error}, advocates that the reported values for shell viscosity are predominantly affected by an approximately constant systematic overestimation of the $\rm FWHM_{\mathcal{H}^2}$.
The elevation of the fitting line gives an estimate of such overestimation, which yields $\Delta f_{\rm err} \approx \SI{270}{\kilo\hertz}$.
The left part of Fig.\ \ref{fig:FigureKappasScalingLaws}(b) compares the shell viscosity values reported by van der Meer \textit{et al.} (green dots) with the shell viscosity estimates (dotted lines) inferred from synthetic microbubble frequency responses [Eq.\ \eqref{eq:LinearOscillatorFourier}] affected by a bandwidth error $\Delta f_{\rm err} = \SI{270}{\kilo\hertz}$.
The black lines correspond to a shell viscosity value $\kappa_s = \SI{1e-8} {\kilo\gram\per\second}$, while the grey lines correspond to a value $\kappa_s = \SI{1e-9} {\kilo\gram\per\second}$.
The excellent agreement found suggests that a systematic overestimation of the $\rm FWHM_{\mathcal{H}^2}$, approximately constant with bubble size, may be the dominant cause for the unphysical trends in the shell viscosity found in several studies.
Figure \ref{fig:FigureKappasScalingLaws}(a) also reveals that the deviation of the shell viscosities found by Daeichin \textit{et al.} follows a quadratic trend in logarithmic scale (dotted red line) which results from the sum of a first contribution that scales with $R_0^3\omega_0\delta$ and a second one that scales with $R_0^3$ (sum in solid red line, single contributions in grey lines), 
unveiling, in view of Eqs.\eqref{eq:viscosity_magnitudespectrum_error} and \eqref{eq:viscosity_bandwidth_error}, that the FWHM is measured from the magnitude instead of the energy of the frequency responses and the bandwidth itself is possibly overestimated, on average, by an amount $\Delta f_{\rm err} \approx \SI{190}{\kilo\hertz}$.
The right part of Fig.\ \ref{fig:FigureKappasScalingLaws}(b) compares the shell viscosity values reported by Daeichin \textit{et al.} (red dots) with the shell viscosity estimates (grey lines) inferred from the \emph{magnitude} of synthetic microbubble frequency responses [Eq.\ \eqref{eq:LinearOscillatorFourier}] affected by a bandwidth error $\Delta f_{\rm err} = \SI{190}{\kilo\hertz}$, yielding again a remarkably good agreement.

The comprehensive specifications reported in the work of van der Meer \textit{et al.} regarding their shell characterisation procedure allow us to directly examine its accuracy by replicating and applying it on synthetic responses of microbubbles of known shell viscosity.
The values of the physical parameters and the specifics of the acoustic driving and of the spectroscopy analysis are taken directly from the information reported in their study.
The model used for the bubble dynamics is nonlinear and based on the modified Rayleigh--Plesset equation [Eq.\ \eqref{eq:RP}] with the Marmottant shell model [Eq.\ \eqref{eq:InterfacePressureMarmottant}] and the polytropic process approximation for the gas contained in the bubble [Eq.\ \eqref{eq:PolyGas}].
The latter allows a smooth carryover of the physical parameters used in the study in the bubble model.
Note that employing the full model presented in \S \ref{BubbleModel} would still lead to the same results that we are going to present. 
The shell viscosity values chosen are again $\kappa_s = \SI{1e-9} {\kilo\gram\per\second}$ and $\kappa_s = \SI{1e-8} {\kilo\gram\per\second}$.

\begin{figure}[!ht] 
    \centering
        \includegraphics[width=\columnwidth]{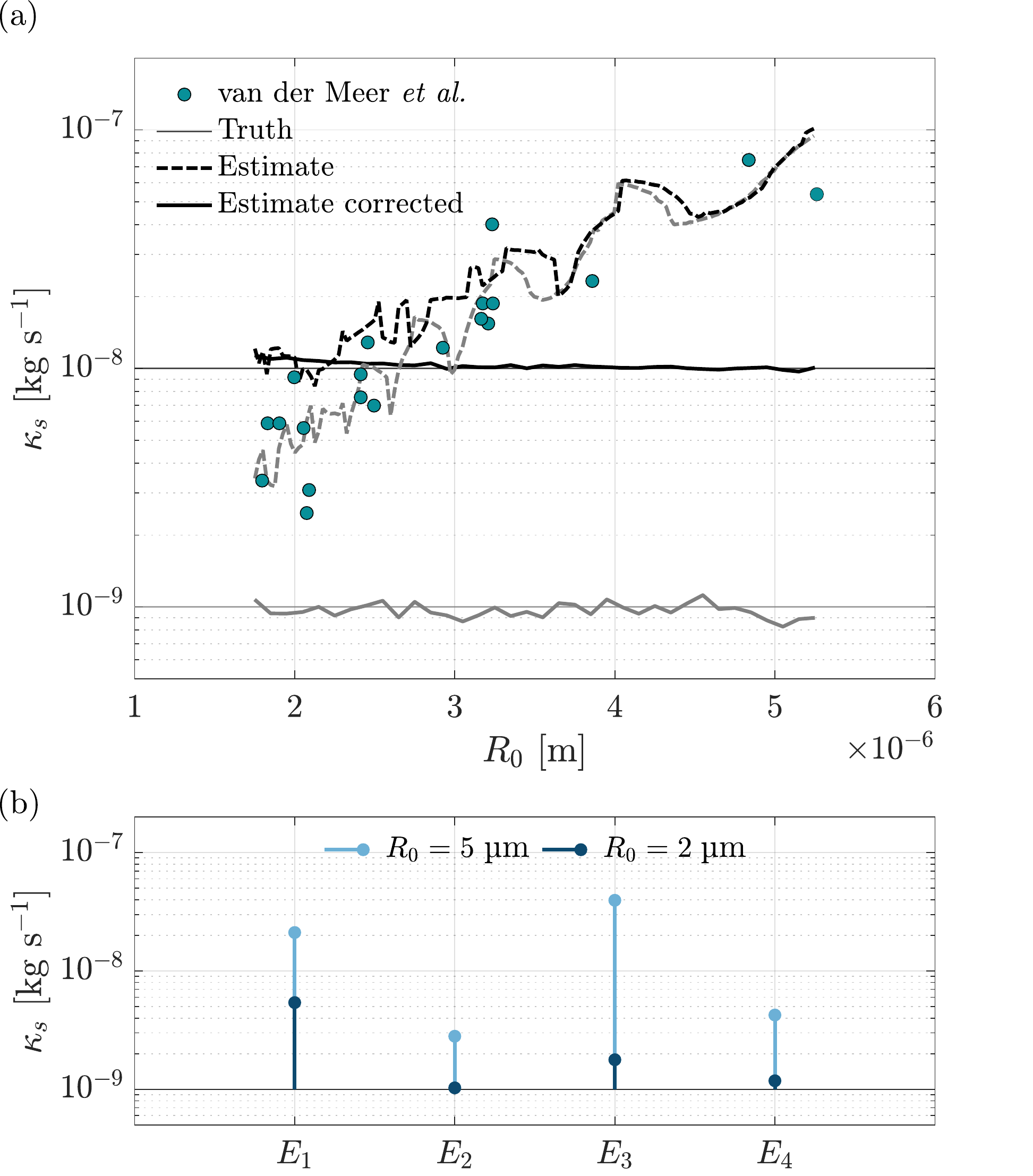}
    \caption{(a) Comparison between the shell viscosity values reported by van der Meer \textit{et al.} \cite{VanDerMeer2007MicrobubbleAgents}, in green dots, and the values estimated applying the spectroscopy methodology detailed in their work to synthetic nonlinear frequency responses of microbubbles of known shell viscosities $\kappa_s = \SI{1e-9} {\kilo\gram\per\second}$ and $\kappa_s = \SI{1e-8} {\kilo\gram\per\second}$, in dashed black and grey lines, respectively. 
    The specifications used, taken from their study, are: $p_a = \SI{40}{\kilo\pascal}, \ \Delta f_{\rm sam} = \SI{50}{\kilo\hertz}, \ T_{\rm obs} = \SI{4.2} {\micro\second}$.
    The solid lines are the result of the same procedure but having reduced the methodological sources of bias using the following specifications: $p_a = \SI{1}{\kilo\pascal}, \ \Delta f_{\rm sam} = \SI{5}{\kilo\hertz}, \ T_{\rm obs} = \SI{100} {\micro\second}$.
    (b) Error breakdown in the estimation of a shell viscosity $\kappa_s = \SI{1e-9} {\kilo\gram\per\second}$ for two different bubble sizes, $R_0 = \SI{2}{\micro\metre}$ and $R_0 = \SI{5}{\micro\metre}$ on the basis of the methodological inaccuracies $E_1-E_4$ present in the work of van der Meer \textit{et al.}.
    Please refer to the text for their description.
    The model of the bubble dynamics used for calculating the frequency responses is based on the modified Rayleigh--Plesset equation [Eq.\ \eqref{eq:RP}] with the Marmottant shell model [Eq.\ \eqref{eq:InterfacePressureMarmottant}] and the polytropic process approximation for the gas contained in the bubble [Eq.\ \eqref{eq:PolyGas}].
    The physical parameters used are: $p_0 = \SI{100} {\kilo\pascal}, \ \sigma_w = \SI{72} {\milli\newton\per\meter}, \ c_l = \SI{1500} {\meter\per\second}, \ \mu_l = \SI{0.002} {\pascal\second}, \ \rho_l = \SI{1000} {\kilo\gram\per\cubic\meter}, \ E_s = \SI{0.54} {\newton\per\meter}, \ \sigma_0 = \SI{0.02} {\newton\per\meter},  \ n = 1.07$.
    }
    \label{fig:Figure_vdM}
\end{figure}

The results of this endeavour are presented in Fig.\ \ref{fig:Figure_vdM}(a) which reveals with the dotted-dashed lines that their methodology leads to an inaccurate estimate of the shell viscosity: its value gets inflated, the scatter for a given bubble size is reduced and a strong dependency on the initial radius appears.
The estimated shell viscosities compare remarkably well with the values reported by the authors (green dots), endorsing the idea that the apparent dependency of the shell viscosity on the bubble equilibrium radius is indeed only the result of an experimental artefact.
The methodological sources of bias occurred in the experimental procedure are diverse: (i) $E_1$: the linear model was applied to bubbles behaving nonlinearly ($p_a = \SI{40}{\kilo\pascal}$); (ii) $E_2$: the frequency response was sampled with too coarse a resolution ($\Delta f_{\rm sam} = \SI{50}{\kilo\hertz}$); (iii) $E_3$: the bubble dynamics was observed for too short a period of time ($T_{\rm obs} = \SI{4.2} {\micro\second} $) leading to spectral leakage in the frequency spectrum of its radial time evolution; (iv) $E_4$: the resonance peak \emph{area} of the Fourier transform of the radial time evolution of the microbubble under investigation was inappropriately considered as the measure of its response at the specified input frequency.
The aftereffect of each source of bias listed above is presented in Fig.\ \ref{fig:Figure_vdM}(b) for a bubble with shell viscosity $\kappa_s = \SI{1e-9} {\kilo\gram\per\second}$ and two different equilibrium radii $R_0 = \SI{2}{\micro\metre}$ and $R_0 = \SI{5}{\micro\metre}$.
Note that the arithmetic sum of these values does not exactly return the results shown in Fig.\ \ref{fig:Figure_vdM}(a) because the errors combine nonlinearly.
To further confirm that the cause of the departure from the prescribed shell viscosity lies in the experimental parameters adopted, the solid thick lines in Fig.\ \ref{fig:Figure_vdM}(a) show that reducing the methodological inaccuracies, detailed above, to negligible values (i.e.\ using $p_a \approx \SI{1}{\kilo\pascal}$, $\Delta f_{\rm sam} < \SI{5}{\kilo\hertz}$, $T_{\rm obs} > \SI{100} {\micro\second} $) the spectroscopy method would be capable of recovering the correct shell viscosity.
Note, however, that the small oscillations resulting from such a low driving pressure would be beyond the resolving capabilities of any existing optical microscopy apparatus.

In this regard, we acknowledge the great potential of the shell investigation technique introduced in the work of Daeichin \textit{et al.}.
The photoacoustic impulsive driving and the detection method based on laser scattering allow one to surmount the limitations due to the use of optical microscopy that the work of van der Meer \textit{et al.} ran up against and, in principle, to obtain very accurate shell viscosity and modulus estimates from linearly oscillating bubbles.
However, we point out that, in the work in question, in addition to having improperly extracted the damping from the half-amplitude instead of half-energy bandwidth, the observation time used is potentially too short and the driving pressure signal might be deviating from an exact Dirac impulse.
These are all factors that can compromise the shell viscosity estimation and possibly explain the reported upward trend with the equilibrium radius.

%
% Figure obtained using "Figure_vdM.m" script

\subsection{Radius-time curve fitting: model and bubble sizing bias affect the shell viscosity estimation}
Finally, we turn our attention to the works conducted by Morgan \textit{et al.}\cite{Morgan2000ExperimentalSize},  Doinikov \textit{et al.} \cite{Doinikov2009ModelingMicrobubbles} and Tu \textit{et al.} \cite{Tu2009EstimatingScattering,Tu2011MicrobubbleCytometry} which estimated the shell viscosity of single microbubbles using an approach very similar to ours: by finding the theoretical radius-time curve that results in the minimum mean square error from the empirical curve, recorded with a streak camera in the first two cases or with a side laser scattering technique in the second two.
These works also found the shell viscosity increasing with the bubble size.
Clearly, the error analysis we carried out for bubble spectroscopy cannot be applied to these studies. However, we believe that their results can also possibly suffer from other methodological oversights causing the upward trend.

 Morgan \textit{et al.}\cite{Morgan2000ExperimentalSize} and  Doinikov \textit{et al.} \cite{Doinikov2009ModelingMicrobubbles}  did not report whether the theoretical model used is able to capture the behaviour of the microbubbles \emph{across a range of sizes}, as we did and shown in Fig.\ \ref{fig:R0VE_RT}(b) and, in particular, whether it can identify the correct resonance radius for the driving frequency employed.
Specifically, the microbubbles were driven at a high pressure amplitude (over $\SI{100}{\kilo\pascal}$) and thus into strong nonlinear oscillations, while the theoretical model employed is based on a linear representation of the coating physics.
Therefore, the resonance peak from the theoretical model is expected to be at a larger radius than that found in the experiments. 
This leads the optimisation algorithm to assign a low viscosity to small-diameter bubbles and vice versa, and thus to an upward trend in the shell viscosity similar to that found in their study.
Tu \textit{et al.} \cite{Tu2009EstimatingScattering,Tu2011MicrobubbleCytometry} did employ a nonlinear model, however, they solely conducted indirect measurements of bubble evolution, by converting scattering light signals. 
This approach raises concerns regarding the reliability of the obtained bubble radius measurements.
It is important to note that any inaccuracies in the radius measurement can introduce distortions in the profile of the bubble behaviour across bubble sizes.
This, in turn, can result in a mismatch between the experimental data and the theoretical model employed for estimating the shell viscosity, significantly impacting the accuracy and reliability of the inferred shell viscosity.
In this regard, Tu \textit{et al.} also did not compare the theoretical model used with the experimental bubble response across bubble sizes.
This omission prevents the resolution of lingering doubts regarding the accuracy of the shell viscosity measurement.

\begin{figure}[!ht] 
    \centering
        \includegraphics[width=\columnwidth]{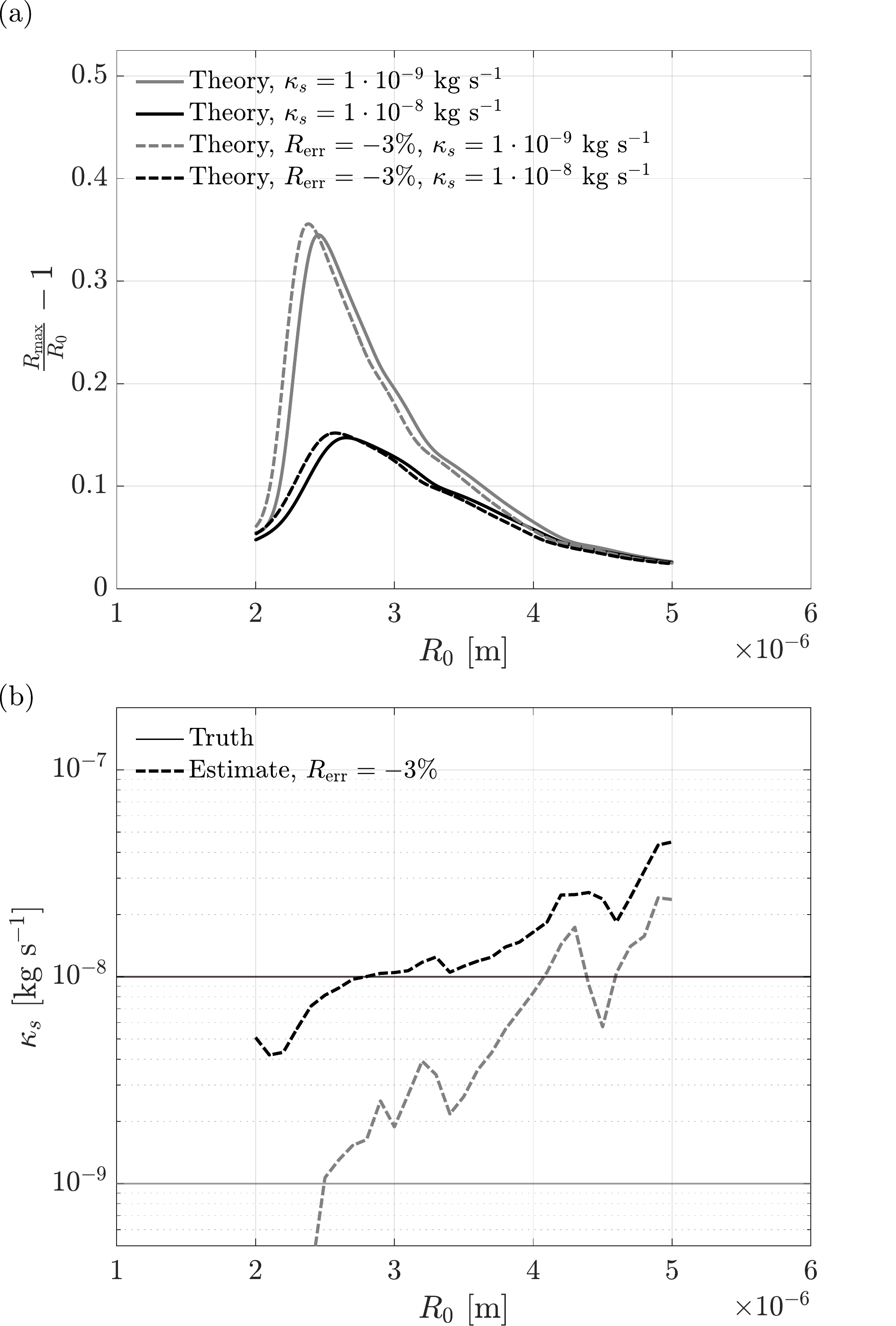}
    \caption{Effectiveness of the radius-time curve fitting in inferring the shell viscosity of microbubbles in the presence of a radius measurement error. 
    (a) The solid lines represent the model  of the bubble dynamics (modified Rayleigh--Plesset equation [Eq.\ \eqref{eq:RP}] with the Marmottant shell model [Eq.\ \eqref{eq:InterfacePressureMarmottant}] and the polytropic process approximation for the gas contained in the bubble [Eq.\ \eqref{eq:PolyGas}]) employed to infer the shell viscosity from radius-time curves of microbubbles.
    The model is presented for equilibrium radii $R_0 = \SIrange{2}{5} {\micro\meter}$ and shell viscosity values $\kappa_s = \SI{1e-9}{\kilo\gram\per\second}$ (grey lines) and $\SI{1e-8} {\kilo\gram\per\second}$ (black lines).
    The dashed lines depict the synthetic dataset generated using the aforementioned model of bubble dynamics but with a deliberate error $R_{\rm err} = -3 \%$ in the radius measurement.
    (b) The solid thin lines represent the shell viscosity to be inferred.
    The dashed lines are the result of the radius-time curve fitting in inferring the shell viscosity of the aforementioned synthetic dataset.
    The physical parameters used are: $f = \SI{1.5} {\mega\hertz}, \ p_a = \SI{40} {\kilo\pascal}, \ p_0 = \SI{100} {\kilo\pascal}, \ \sigma_w = \SI{72} {\milli\newton\per\meter}, \ c_l = \SI{1500} {\meter\per\second}, \ \mu_l = \SI{0.002} {\pascal\second}, \ \rho_l = \SI{1000} {\kilo\gram\per\cubic\meter}, \ E_s = \SI{0.54} {\newton\per\meter}, \ \sigma_0 = \SI{0.02} {\newton\per\meter},  \ n = 1.07$.}
    \label{fig:RadiusMeasurementErrors}
\end{figure}

To gain a better understanding of the sensitivity of the shell viscosity estimate on the radius measurement error, let us consider synthetic data of the response of bubbles across different sizes [represented by dashed lines in Fig.\ \ref{fig:RadiusMeasurementErrors}(a)], generated using a model of bubble behaviour [illustrated by solid lines in Fig.\ \ref{fig:RadiusMeasurementErrors}(a)], but with a deliberate error $R_{\rm err} = -3 \%$ introduced in the radius measurement.
If the aforementioned model is subsequently employed to infer the shell viscosity, the error induced by the inaccurate bubble sizing will propagate to the shell viscosity estimate, leading to clear trends correlating with the bubble radius, as shown in Fig.\ \ref{fig:RadiusMeasurementErrors}(b).
This arises from the discrepancy between the positions of the resonance peaks in the model and the synthetic data.
It is evident that if the error on the radius has an opposite sign, the trends will exhibit a reversal in their slope.
To avoid these undesirable effects in our work, we carefully verified the agreement between the model and the experimental data, and thus the absence of errors on either side, prior to estimating shell viscosity, as portrayed in Fig.\ \ref{fig:R0VE_RT}(b).

\section{Conclusions}
In this work, we conducted measurements of the dilatational viscosity of the shell of ultrasound contrast agent microbubbles.
By combining ultra-high-speed microscopy imaging, optical trapping and wide-field fluorescence, we obtained recordings of the response of individual microbubbles to an ultrasound driving pulse with unprecedented accuracy.
We validated an advanced theoretical model of bubble dynamics based on the compressible Rayleigh--Plesset equation, and augmented by the Marmottant  model to account for the shell and the Zhou model to accurately track the pressure variation in the bubble, against the experimental data collected, and found excellent agreement, both with the individual bubble responses and the overall response curve across the entire range of bubble sizes.
This model was therefore employed in an optimisation routine to solve the inverse problem of determining the shell viscosity of single bubbles from their radial time evolution.
The resulting shell viscosity values show a large, one order of magnitude variability, ranging approximately from $\kappa_s = \SI{1e-9} {\kilo\gram\per\second}$ to $\kappa_s = \SI{1e-8} {\kilo\gram\per\second}$, and an average value of $\overline{ \kappa_s} = \SI{5.4e-9} {\kilo\gram\per\second}$.
Most interestingly, we found no dependency of shell viscosity on the bubble equilibrium radius.
These results contrast with more than two decades of previous studies, which have reported less variability and, very intriguingly, a strong upward trend as the bubble equilibrium radius increases.
The latter behaviour is considered to be unphysical, nevertheless, it has thus far never been either explained or proven as such.
We believe that the root cause of this issue in the majority of studies lies in the high sensitivity to errors of "bubble spectroscopy", the method employed to extract shell viscosity from the bubble frequency response bandwidth.
We demonstrated with analytical arguments that an overestimation of the bandwidth leads to inflated shell viscosity values, reduced dispersion for a given bubble size, and the appearance of a strong dependence on the equilibrium radius. 
We pinpointed several methodological inaccuracies in previous studies that lead to an overestimation of bandwidth, including: the use of linear models to describe the nonlinear bubble response, a too short observation time of the bubble response, a too low frequency sampling rate, the use of the half-magnitude bandwidth instead of the half-energy bandwidth, to name a few.
We exemplified that resolving these inaccuracies allows the same trend found in our measurements to be recovered.
Furthermore, our findings indicate that the unphysical trends observed in the remaining studies, where the shell viscosity was determined by fitting radius-time curves, may be attributed to errors in bubble sizing. This issue leads to a misalignment of resonance peaks between the experimental data and the bubble dynamics model used to infer the shell viscosity. 
As a consequence, the optimisation procedure assigns incorrect shell viscosity values to the microbubbles, resulting in trends that falsely correlate with the bubble radius.
Finally, we suggested that the variability found in the shell viscosity values reflects the arbitrariness in the size and distribution of the phospholipid domains among the bubble shells.
The results presented could provide useful guidelines for characterising the shell properties of microbubbles which is critical for predicting their bioeffects in medical applications.

\section*{Author Contributions}
\textbf{Marco Cattaneo}: Conceptualisation, Methodology, Software, Formal analysis, Investigation, Data curation, Writing - Original Draft, Visualisation, Project administration. 
\textbf{Outi Supponen}: Conceptualisation, Methodology, Investigation, Writing - Review \& Editing, Project administration, Resources, Supervision.
\section*{Conflicts of interest}
There are no conflicts to declare.

\section*{Acknowledgements}
We thank Dr. Gazendra Shakya for the preparation of microbubbles. We also thank Prof. Jan Vermant for insightful discussions about phospholipid monolayer rheology. We acknowledge ETH Zürich for the financial support.

\bibliography{output}

\end{document}